\documentclass[12pt,aps,nofootinbib,preprintnumbers,groupedaddress,letterpaper]{article}

\usepackage{hyperref}
\usepackage[dvipsnames]{xcolor}
\usepackage[normalem]{ulem}
\usepackage{amsmath, bbold}
\usepackage{enumerate}
\usepackage{amsfonts}
\usepackage{yfonts}

\usepackage{graphicx}

\newcommand\comment[1]{}

\newcommand\poincare{Poincar\' e }

\newcommand\ov{\over }

\def\le{\left}
\def\ri{\right}

\def\({\left(}
\def\){\right)}
\def\<{\langle}
\def\>{\rangle}

\def\tr{\mathop{\rm tr}}

\newcommand\half{{\ensuremath{\frac{1}{2}}}}
\newcommand\p{\ensuremath{\partial}}

\newcommand\field[1]{{\ensuremath{\mathbb{{#1}}}}}

\newcommand{\CC}{\field{C}}

\newcommand{\RR}{\field{R}}

\newcommand{\be}{\begin{equation}}
\newcommand{\ee}{\end{equation}}
\newcommand{\bea}{\begin{eqnarray}}
\newcommand{\eea}{\end{eqnarray}}
\newcommand{\bwt}{\begin{widetext}}
\newcommand{\ewt}{\end{widetext}}

\newcommand{\bi}{\begin{itemize}}
\newcommand{\ei}{\end{itemize}}
\newcommand{\ben}{\begin{enumerate}}
\newcommand{\een}{\end{enumerate}}
\newcommand{\bca}{\begin{cases}}
\newcommand{\eca}{\end{cases}}
\newcommand{\bln}{\begin{align}}
\newcommand{\eln}{\end{align}}
\newcommand{\bst}{\begin{split}}
\newcommand{\est}{\end{split}}

\begin{document}

\begin{titlepage}

\begin{flushright}
QMUL-PH-21-28\\
\end{flushright}

\vspace{5mm}

  \begin{center}

\centerline{\Large \bf {The spectral curve of segmented strings}}

\bigskip
\bigskip

{\bf David Vegh}

\bigskip

\small{
{ \it  Centre for Research in String Theory, School of Physics and Astronomy \\
Queen Mary University of London, 327 Mile End Road, London E1 4NS, UK}}

\medskip

{\it email:} \texttt{d.vegh@qmul.ac.uk}

\medskip

{\it \today}

\bigskip


\begin{abstract}

I show how to compute the spectral curve of piecewise linear Nambu-Goto strings in  three-dimensional anti-de Sitter spacetime in terms of  `celestial' embedding variables.

\end{abstract}

\end{center}

\end{titlepage}

\tableofcontents

\clearpage

\section{Introduction}

This paper is concerned with the classical bosonic string in three-dimensional anti-de Sitter (AdS) spacetime. The Nambu-Goto action is simply the area of the worldsheet and thus solutions to the equation of motion are extremal surfaces in spacetime.
There are several reasons to study this system. String theory in asymptotically AdS spacetimes is dual to strongly coupled gauge theories \cite{Maldacena:1997re, Gubser:1998bc, Witten:1998qj}.
On both sides of AdS/CFT the theories are in a certain sense integrable which allows for non-trivial tests of the correspondence (see \cite{Bena:2003wd, Minahan:2002ve} and the review \cite{Beisert:2010jr} and references therein). The equation of motion of  a long string in AdS$_3$ is equivalent to the generalized sinh-Gordon model. Simple solutions correspond to rotating folded strings \cite{Gubser:2002tv}. Excited states with (singular) sinh-Gordon solitons correspond to cusps (spikes) on the string \cite{Gubser:2002tv, Kruczenski:2004wg, Jevicki:2007aa, Jevicki:2008mm, Dorey:2008vp, Jevicki:2009uz, Dorey:2010iy, Dorey:2010id, Irrgang:2012xb}.

Another motivation to study the system comes from the realization that the worldsheet theory on a long string behaves like a two-dimensional toy model for gravity. Although the worldsheet theory has neither black holes nor the usual graviton excitations, it shares some features with theories of quantum gravity \cite{Dubovsky:2012wk}. At finite temperature one can see the saturation of the `chaos bound' \cite{Maldacena:2015waa} and the emergence of chaos on the string \cite{deBoer:2017xdk}. For near-AdS$_2$ embeddings the  Schwarzian action  \cite{kitaev, Maldacena:2016hyu, Jensen:2016pah, Maldacena:2016upp, Engelsoy:2016xyb} can be derived \cite{Banerjee:2018twd, Vegh:2019any}.
Finally, we note that the system also provides a controlled laboratory for studying non-linear phenomena such as wave-turbulence and energy cascades \cite{Ishii:2015wua, Vegh:2018dda}. This is made possible by the existence of exact methods which allows one to perform numerical calculations without accumulating errors.

In this paper, instead of studying the smooth string (possibly with a finite number of cusps present), we investigate the  discretized version of the equation of motion. The discretization is exact, i.e. it preserves integrability and solutions do extremize the Nambu-Goto action. The discrete equation of motion is arguably simpler than the  partial differential equation (which is supplemented with constraints). The corresponding embeddings are piecewise linear (a.k.a. segmented) strings. The goal of the paper is to compute a basic invariant: the {\it spectral curve} for segmented strings. This can be defined by computing the monodromy of the Lax connection which will be the topic of the next section. Section 3 gives an introduction to segmented strings. Section 4 solves the forward scattering problem on segmented strings, then section 5 computes the Lax matrix in terms of {\it celestial} variables. Section 6 discusses the properties of the spectral curve for closed strings. A simple example with four segments is presented in detail in Section 7. The paper ends with a summary of the results and with a discussion of possible future research directions.

\clearpage

\section{Strings in AdS$_3$}

In this section we collect the basic equations and quantities that describe the Nambu-Goto string in AdS$_3$.
A unit size AdS$_3$ can be immersed into an $\RR^{2,2}$ ambient space via
\be
  \nonumber
   Y \cdot  Y \equiv -Y_{-1}^2 - Y_0^2 + Y_1^2 + Y_2^2 = -1 \ , \qquad  Y \in \RR^{2,2}.
\ee
The boundary of AdS is the set of points that satisfy $ Y^2 = 0$ with the identification $ Y \cong a  Y$ (with $a \in \RR^+$).
Classically, the Nambu-Goto action is equivalent to the sigma model action
\be
   \nonumber
 S  = -{T \ov 2}\int d\tau d\sigma ( \p_\sigma Y^\mu \p_\sigma Y_\mu - \p_\tau Y^\mu \p_\tau Y_\mu + \lambda( Y^2 + 1)) ,
\ee
where $T$ is the string tension and $ Y(\tau, \sigma) \in \RR^{2,2}$ is the embedding function. The string is mapped into $\RR^{2,2}$ and the Lagrange multiplier $\lambda$ forces the string to lie on the hyperboloid.
In lightcone coordinates
\be
  \nonumber
  z = \half(\tau-\sigma) \ , \quad \bar z = \half(\tau+\sigma)  \ , \quad  \p \equiv \p_z = \p_\tau - \p_\sigma  \ , \quad \bar\p \equiv \p_{\bar z} = \p_\tau + \p_\sigma
\ee
the equation of motion is given by
\be
   \nonumber
 \label{eq:eoms}
  \p \bar\p Y - (\p Y \cdot \bar\p Y )  Y = 0 \, .
\ee
Due to the gauge choice, the equations are supplemented by the Virasoro constraints
\be
  \label{eq:virasoro}
  \p Y \cdot \p Y = \bar\p Y \cdot \bar\p Y = 0 \, .
\ee

\subsection{The generalized sinh-Gordon model}

Let us start by defining the {\it sinh-Gordon field} which can be computed from the embedding via
\be
   \nonumber
 e^{2\alpha(z, \bar z)} = \half \p Y \cdot \bar\p Y  \, .
\ee
A {\it normal vector} to the worldsheet is given by
\be
  \nonumber
    N_a  = \half e^{-2\alpha} \epsilon_{abcd} Y^b \p Y^c \bar\p Y^d
\ee
which satisfies $N\cdot Y = N \cdot \p Y =N \cdot \bar\p Y = 0$ and $N^2 = 1$.
Finally, let us define the following {\it auxiliary fields}
\bea
  \nonumber
  2u &=& \p N \cdot\p Y = -N\cdot \p\p Y  =  - Y \cdot \p\p N \, \\
  \label{eq:discv}
  2v &=&  -\bar\p N \cdot\bar\p Y = N\cdot \bar\p\bar\p Y  =  Y \cdot \bar\p\bar\p N \ .
\eea
The string equations of motion imply that $u = u(z)$ and $v=v(\bar z)$ and that the quantities satisfy the generalized sinh-Gordon equation \cite{Pohlmeyer:1975nb, DeVega:1992xc}
\be
   \label{eq:sg}
   \p \bar\p \alpha - e^{2\alpha} + uv e^{-2\alpha} = 0 \ .
\ee

\subsection{Lax connection}

Let us  consider the following matrix built from the embedding vector, the normal vector, and the sinh-Gordon field \cite{Alday:2009yn}
\be
  \label{eq:wmat}
  W= \half \begin{pmatrix}
Y+N & {e^{ {-\alpha }}\bar\p Y  }\cr
 {e^{-  {\alpha }}  }\p Y  & Y-N
\end{pmatrix} \, .
\ee
Using the equivalence of $SO(2,2)$ and $SL(2)\times SL(2)$, spacetime indices can be decomposed into spinor indices, e.g.
\be
  \label{eq:yeq}
 Y^\mu \to Y_{a \dot a } =
\begin{pmatrix}
Y_{-1}+Y_{2}& Y_1-Y_0\cr Y_1 + Y_0 & Y_{-1}-Y_{2}
\end{pmatrix}_{a\dot{a}} \, .
\ee
Elements of the matrix will be denoted by $W_{\alpha\dot \alpha, a\dot a}$ where $\alpha \in \{0,1\}$ and $\dot \alpha \in \{\dot{0},\dot{1}\}$ denote the rows and columns of (\ref{eq:wmat}) while $a\in \{0,1\}$ and $\dot a \in \{\dot{0},\dot{1}\}$ denote spacetime indices.
The string equations of motion can be written as
\bea
  \nonumber
  & \p W_{\alpha\dot \alpha, a\dot a}+(B_z^L)_{\alpha}^{~\beta}W_{\beta\dot \alpha, a\dot a}
  + (B_z^R)_{\dot \alpha}^{~\dot \beta} W_{\alpha\dot \beta, a\dot a} = 0 & \\
  \nonumber
  & \bar\p W_{\alpha\dot \alpha, a\dot a}+(B_{\bar z}^L)_{\alpha}^{~\beta}W_{\beta\dot \alpha, a\dot a}
  + (B_{\bar z}^R)_{\dot \alpha}^{~\dot \beta} W_{\alpha\dot \beta, a\dot a} = 0 &
\eea
with left (L) and right (R) $SL(2)$ Lax connections given by
\bea
\nonumber
B_z^L=\begin{pmatrix}
{1 \over 2} \p \alpha & {-e^{ {\alpha }}  }\cr
-{u(z)}{e^{-  {\alpha }}  }  & -{1 \over 2} \p \alpha
\end{pmatrix} \, , &&
B_{\bar{z}}^L=\begin{pmatrix}
-{1 \over 2} \bar\p \alpha & -{v(\bar z) }{e^{-{\alpha } }   }\cr
-e^{ {\alpha }}    & {1 \over 2} \bar\p \alpha
\end{pmatrix}   \, , \\
\nonumber
B_z^R=\begin{pmatrix}
-{1 \over 2} \p \alpha &{{u(z)} e^{- {\alpha }}   }   \cr
-{e^{ {\alpha }}  } & {1 \over 2} \p \alpha
\end{pmatrix}  \, ,
&&
B_{\bar{z}}^R=\begin{pmatrix}{1 \over 2} \bar\p \alpha & -{e^{ {\alpha}} }  \cr
{v(\bar z)}{e^{- {\alpha}}   }  & -{1 \over 2} \bar\p \alpha \end{pmatrix} \, .
\eea
It is easy to check that consistency of the above equations imply the flatness conditions
\be
\label{eq:flatness}
\partial B_{\bar z}^{L}-\bar{\partial}B_z^{L}+[B_z^{L},B_{\bar{z}}^{L}] = 0   \, , \quad
\partial B_{\bar z}^{R}-\bar{\partial}B_z^{R}+[B_z^{R},B_{\bar{z}}^{R}] = 0
\ee
which in turn are equivalent to the sinh-Gordon equation (\ref{eq:sg}).

Given a solution of the generalized sinh-Gordon equation, one would like to compute the corresponding string embedding. In order to do this one has to solve the following auxilliary scattering problem.
Let us consider the  linear systems
\bea
\nonumber
\p \psi^L_{\alpha}+(B_z^L)_{\alpha}^{~\beta}\psi^L_{\beta}=0 \, , & &
\bar\p \psi^L_{\alpha}+(B_{\bar{z}}^L)_{\alpha}^{~\beta}\psi^L_{\beta}=0 \, , \\
\label{eq:linear}
\p \psi^R_{\dot \alpha }+(B_z^R)_{\dot \alpha}^{~\dot \beta}\psi^R_{\dot \beta } =0 \, , & &
\bar\p \psi^R_{\dot \alpha }+(B_{\bar{z}}^R)_{\dot \alpha}^{~\dot \beta}\psi^R_{\dot \beta } =0 \, ,\quad
\eea
where $\psi^L_{\alpha}$ and $\psi^R_{\dot \alpha }$ are two-component spinors ($\alpha, \dot \alpha \in \{ 0,1 \}$). Each of these systems have two linearly independent solutions, denoted by $\psi^L_{\alpha a}$   and $\psi^R_{\dot \alpha \dot a}$. Here $ a, \dot  a \in \{ 0,1 \}$ label the independent solutions which are normalized so that
\be
  \label{eq:normalize}
  \epsilon^{ \alpha \beta} \psi^L_{\alpha a} \psi^L_{\beta b}=\epsilon_{a b},~~~~~\epsilon^{\dot \alpha \dot \beta } \psi^R_{\dot \alpha  \dot a} \psi^R_{\dot \beta \dot b}=\epsilon_{\dot a \dot b}
\ee
where $\epsilon$ is the $2\times 2$ Levi-Civita tensor.
The tensor $W$ can be written as
\be
  \nonumber
  W_{\alpha\dot \alpha, a\dot a} =  \psi^L_{\alpha a}\psi^R_{\dot \alpha \dot a} \, .
\ee
Finally, the trace of (\ref{eq:wmat}) is equal to $Y$ and thus the string embedding is given by the expression
\be
  \label{eq:physical}
 Y_{a \dot a }  =
\psi^L_{\alpha a} M_1^{\alpha \dot \beta} \psi^R_{\dot \beta \dot{a}} \, , \qquad
M_1^{\alpha \dot \beta}=\begin{pmatrix}
1 & 0 \cr
0  & 1
\end{pmatrix} \, .
\ee

\subsection{Spectral parameter}
In order to define the spectral curve, let us introduce a {\it spectral parameter} $\zeta \in \CC$ via
\be
\nonumber
B_z=\begin{pmatrix}
{1 \over 2} \p \alpha & {-{1 \ov \zeta}e^{ {\alpha }}  }\cr
-{u \ov \zeta}{e^{-  {\alpha }}  }  & -{1 \over 2} \p \alpha
\end{pmatrix} \, , \quad
B_{\bar{z}}=\begin{pmatrix}
-{1 \over 2} \bar\p \alpha & -{\zeta v }{e^{-{\alpha } }   }\cr
-\zeta e^{ {\alpha }}    & {1 \over 2} \bar\p \alpha
\end{pmatrix}   \, .
\ee
The flatness conditions (\ref{eq:flatness}) are still satisfied for any value of $\zeta$.
Note that the left and right connections can be obtained at special values of the spectral parameter \cite{Alday:2009yn},
\be
  \label{eq:special}
  B_z^L=B_z(\zeta = 1) \, , \quad  B_z^R=U  B_z(\zeta = i) U^{-1} \, , \quad U=\begin{pmatrix}
0 & e^{i\pi/4} \cr
e^{i3\pi/4}  & 0
\end{pmatrix} \, .
\ee
Solutions to the linear problem
\be
\label{eq:linear}
\p \psi_{\alpha}+(B_z)_{\alpha}^{~\beta}\psi^L_{\beta}=0 \, ,  \quad
\bar\p \psi_{\alpha}+(B_{\bar{z}})_{\alpha}^{~\beta}\psi_{\beta}=0
\ee
suffer an $ SL(2)$ monodromy around the loop of a closed string. Our goal is to compute this monodromy for so-called segmented strings which will be defined in the next section.

\clearpage

\begin{figure}[h]
\begin{center}
\includegraphics[width=5cm]{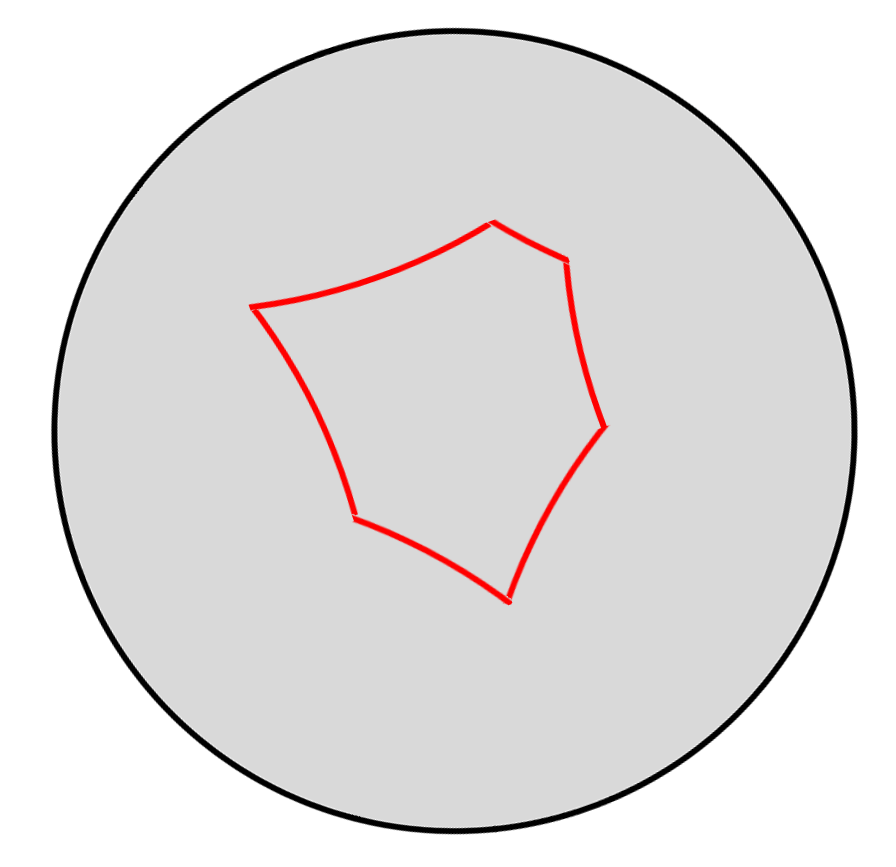} \qquad\qquad
\includegraphics[width=6cm]{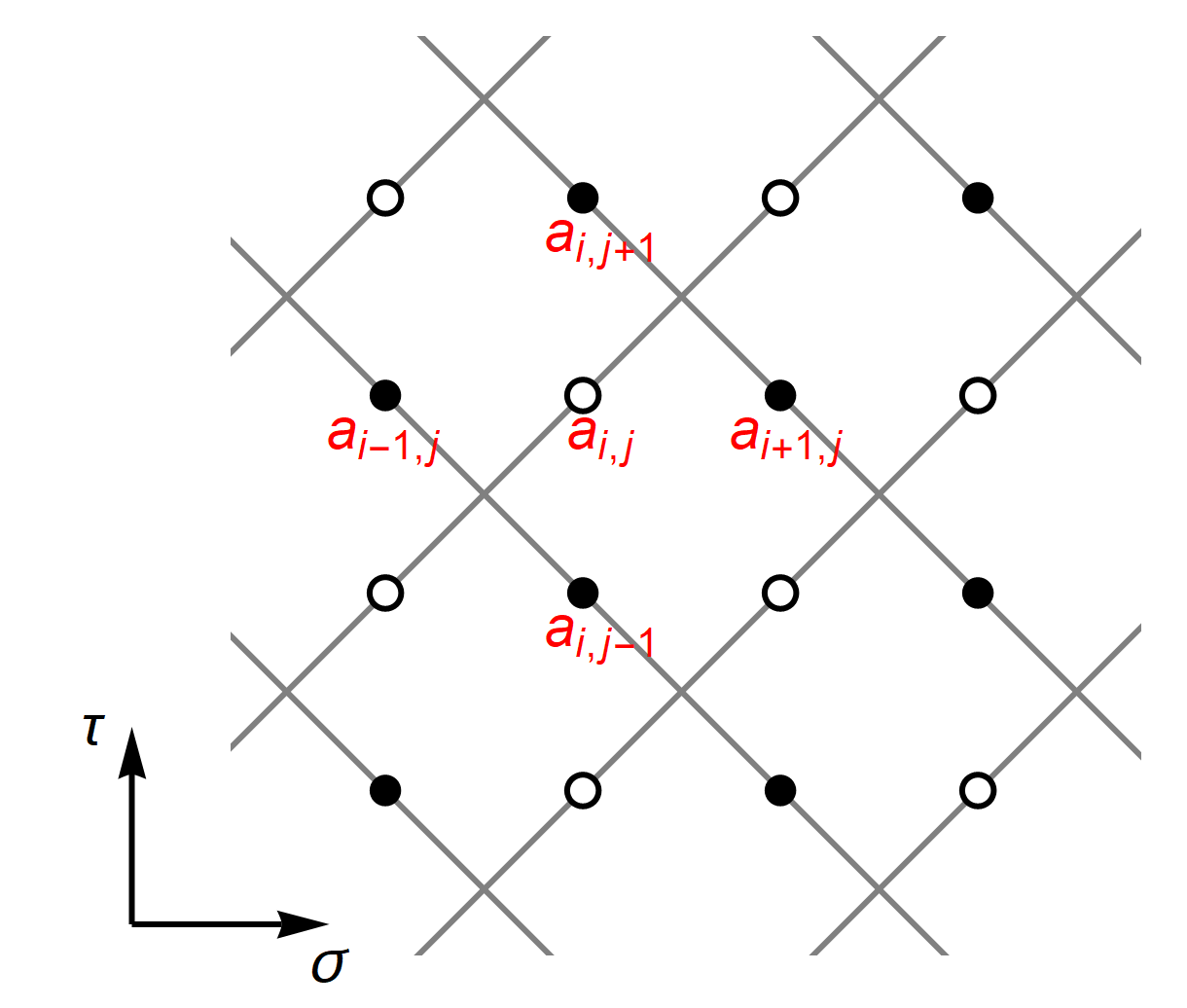}
\caption{\label{fig:sub} {\it Left:} Segmented string on a timeslice of global AdS$_3$. The individual segments are linear AdS$_2$ subspaces in $\RR^{2,2}$. In the example there are six kinks moving with the speed of light. {\it Right:}  The worldsheet of segmented strings. Kink worldlines form a rectangular lattice on the worldsheet. Their motion is characterized by constant null tangent vectors $X \in \RR^{2,2}$ which only change when they collide with other kinks. The `celestial' field $a= {X_{-1} + X_2 \over X_0 + X_1}$ lives on the edges (indicated by black and white dots depending on edge orientation).
}
\end{center}
\end{figure}

\section{Segmented strings}

The aim of this paper is to compute the spectral curve of strings in AdS$_3$. Instead of considering generic strings, we will restrict our attention to segmented strings which form a dense set in the space of all string solutions. Segmented strings are appropriate generalizations of piecewise linear strings in Minkowski spacetime: individual string segments are linear subspaces in $\RR^{2,2}$ \cite{Vegh:2015ska, Callebaut:2015fsa}. For a closed string example on a timeslice of global AdS$_3$, see Figure \ref{fig:sub} (left). The string in the example (red line) consists of six segments connected by {\it kinks}. Each of the segments have a constant normal vector. Kinks must move with the speed of light\footnote{Otherwise in the rest frame of a kink it is easy to see that the string would immediately lose the piecewise linear property.} which implies that the scalar product of normal vectors of adjacent patches must be equal to one.
Due to the  Virasoro constraints (\ref{eq:virasoro}), kinks also move on null lines on the worldsheet. Hence their worldlines form a rectangular lattice as seen in Figure \ref{fig:sub} (right).
Each diamond in the figure is a patch of AdS$_2$ with a constant normal vector. Whenever two kinks collide, a string segment vanishes. The collision results in a new segment with a different normal vector which can be computed by a reflection formula. Time evolution can also be obtained by an alternative reflection formula which computes the fourth kink collision spacetime point from the other three on any AdS$_2$ diamond on the woldsheet \cite{Vegh:2015ska, Callebaut:2015fsa, Vegh:2016hwq, Gubser:2016wno, Vegh:2016fcm}.

\subsection{Celestial variables}

In \cite{Vegh:2016hwq} I showed that segmented strings can be described by a discrete-time affine Toda model. Kink worldlines have constant null tangent vectors which can be decomposed into products of spinor pairs. The area of the worldsheet can be expressed in terms of these spinors which results in the equation of motion
\be
  \label{eq:deqn}
  \hskip -0.15cm {1\ov a_{ij} - a_{i,j+1}}+   {1\ov a_{ij} - a_{i,j-1}} =
  {1\ov a_{ij} - a_{i+1,j}}+   {1\ov a_{ij} - a_{i-1,j}} \, .
\ee
Here $i$ and $j$ are integer indices labeling kink worldlines as shown in Figure \ref{fig:sub} (right). Black and white dots indicate the orientation of the edges. The field $a_{ij}\in \RR$ is the {\it celestial variable} which can be computed by the formula
\be
  \label{eq:aeq}
  {a} = {X_{-1} + X_2 \over X_0 + X_1}  \, ,
\ee
where $X$ is the null tangent vector of the kink. Hence, this field lives on the edges on the kink lattice. Null vectors can be rescaled by a constant factor but this ambiguity drops out of the expression. The equation (\ref{eq:deqn}) is non-linear and produces interesting chaotic phenomena \cite{Vegh:2018dda}. The Cauchy data can be given by specifying two rows of celestial variables, e.g. $a_{i,0}$ and $a_{i,1}$ for all $i$. The variables have to satisfy certain inequality constraints to make sure that the patch areas are non-negative. The string embedding can be obtained from the celestial variables up to a global $SL(2)$ isometry transformation \cite{Vegh:2016hwq}.

In the continuum limit the discrete $a$ field becomes two separate real scalar fields: $b(z, \bar z)$ and $w(z, \bar z)$ named after the color of the dots.
These fields are $\RR^{2,2}$ analogs of coordinates on the  {\it celestial sphere} which is the sphere at null infinity in Minkowski spacetime (see \cite{Pasterski:2016qvg} and also \cite{Atanasov:2021oyu} for the scattering problem in $(2,2)$ signature).

The celestial fields satisfy the equation of motion\footnote{Note that these equations constitute the Lorentzian version of the equation of motion of `stringy cosmic strings'
\be
  \nonumber
  \p \bar \p \tau + {2 \p \tau \bar \p \tau \ov \bar \tau - \tau } = 0
\ee
where  $\tau(z,\bar z)$ is the complex structure modulus of the torus fiber and $z = x_1 + i x_2$ is the two-dimensional Euclidean base \cite{Greene:1989ya}.} \cite{Vegh:2019any}
\be
  \label{eq:eomblack}
   \p \bar \p b = {2 \p b \bar \p b \over b-w } \, ,
\qquad
  \p \bar \p w = {2 \p w \bar \p w \over w-b } \, ,
\ee
which can be derived from either of the following actions
\be
  \label{eq:bwaction}
  S_1 =   2 \int {\bar\p b \p w  \over (b-w)^2}  dz d\bar z \, , \qquad
  S_2 =  2 \int {\bar\p w \p b  \over (b-w)^2}  dz d\bar z \, .
\ee
Henceforth we will use $b$ and $w$ instead of $a$ to denote the celestial field even in the discrete case. The null tangent vector satisfies either $X \propto \p Y$ or $X\propto \bar\p Y$ depending on which way the kink moves.
From (\ref{eq:yeq}) one can see that the celestial field can be computed by taking the ratio of the two elements in the first column.
Therefore the $b$ and $w$ fields can be computed from the $W$ tensor (\ref{eq:wmat}) by \cite{Vegh:2019any}
\be
  \label{eq:bdef}
 b(z, \bar z)  = {(\p Y)_{-1}+ (\p Y)_{2} \over (\p Y)_{0}+ (\p Y)_{1}} = {\psi^L_{10} \psi^R_{\dot{0}\dot{0}}  \ov \psi^L_{11} \psi^R_{\dot{0}\dot{0}} }= {\psi^L_{10}    \ov \psi^L_{11}  } \, ,
\ee
where the first (second) spinor index is the internal (spacetime) index. Similarly we have
\be
  \label{eq:wdef}
    w(z, \bar z)  = {(\bar\p Y)_{-1}+ (\bar\p Y)_{2}  \over (\bar\p Y)_{0}+ (\bar\p Y)_{1}} =  {\psi^L_{00}    \ov \psi^L_{01}  }
\ee
Note that the formulas only depend on the spinors of the left system.

\clearpage

\begin{figure}[h]
\begin{center}
\includegraphics[width=3.7cm]{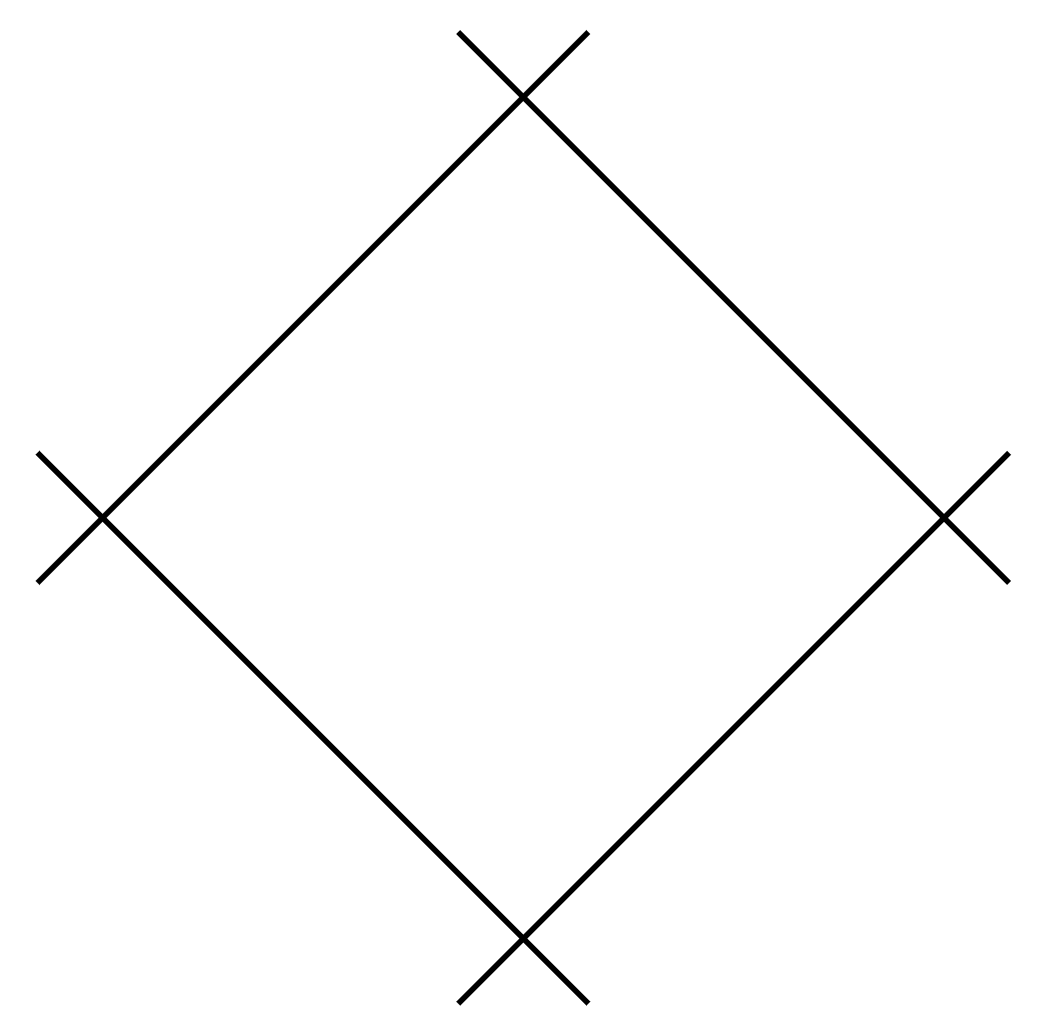} \qquad\qquad
\includegraphics[width=5.8cm]{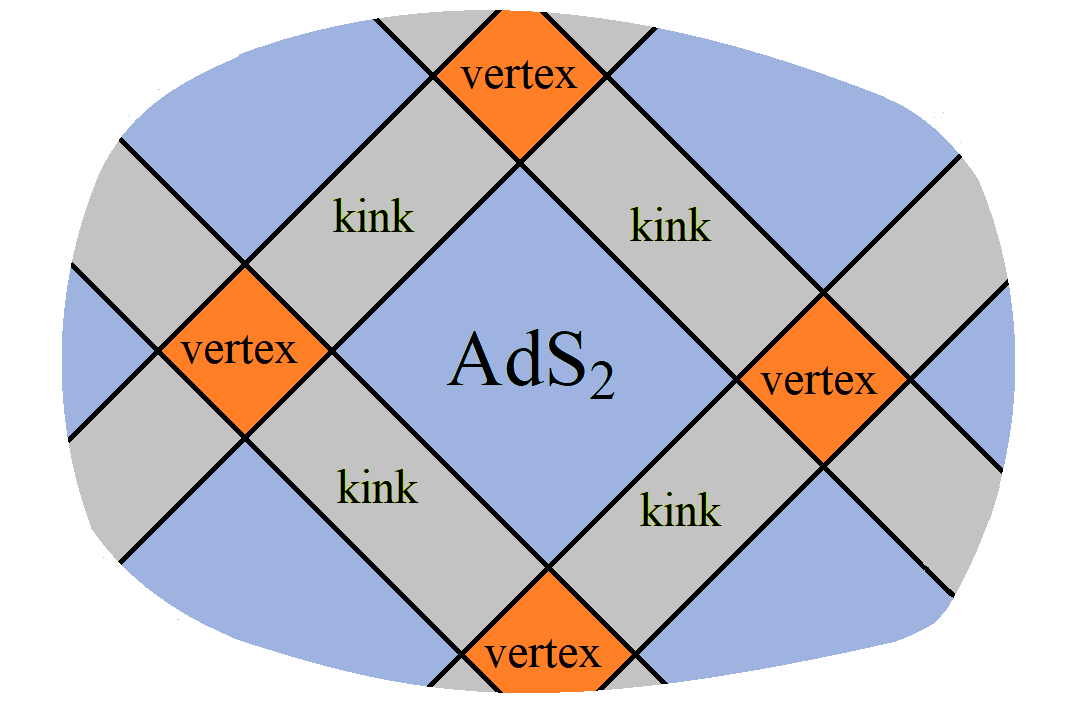}
\caption{\label{fig:wsheet0}  {\it Left:}  The vicinity of an AdS$_2$ patch on the worldsheet. The patch is bounded by the worldlines of four kinks. {\it Right:}  The same worldsheet patch with rescaled coordinates which blow up the vertex (orange) and kink (gray) regions.  In the AdS$_2$ regions  (blue), the normal vector is constant and the position vector changes. In the vertex regions, the position vector is constant and the normal vector changes continuously.
}
\end{center}
\end{figure}

\section{The scattering problem}

In order to compute the Lax monodromy, we need to find solutions to the linear problem in eqn. (\ref{eq:linear}). For segmented strings this will be facilitated by the fact that on different patches the equations simplify and explicit solutions can be found.

Figure \ref{fig:wsheet0} (left) shows the vicinity of an AdS$_2$ patch on the worldsheet. Black lines indicate kink worldlines. Inside the diamond the area density is equal to $2e^{2\alpha(z, \bar z)}$ which is a finite quantity. On the other hand, $u$ and $v$ are vanishingly small and thus $\alpha$ satisfies the Liouville equation. At the kink collision vertices, $u$ and $v$ blow up. By a conformal transformation this divergence can be eliminated. The new coordinate system is shown in Figure \ref{fig:wsheet0} (right). The vertex regions are blown up into orange diamonds. In these regions the area density vanishes while the {\it dual area density} $\beta$ remains finite. It can be defined by
\be
  \nonumber
  e^{2\beta(z, \bar z)} = \half \p N \cdot \bar\p N  \, ,
\ee
where $N$ is the normal vector.

In the following, we will compute the spinor solutions in the vertex and AdS$_2$ regions and then we will match them in the overlap region.

\subsection{AdS$_2$ patches}
In AdS$_2$ regions we have $u=v=0$ and thus the Lax connection simplifies to
\be
\nonumber
B_z=\begin{pmatrix}
{1 \over 2} \p \alpha & {-{1 \ov \zeta}e^{ {\alpha }}  }\cr
0  & -{1 \over 2} \p \alpha
\end{pmatrix} \, , \quad
B_{\bar{z}}=\begin{pmatrix}
-{1 \over 2} \bar\p \alpha & 0 \cr
-\zeta e^{ {\alpha }}    & {1 \over 2} \bar\p \alpha
\end{pmatrix}   \, ,
\ee
and the generalized sinh-Gordon field  solves the Liouville equation $\p \bar\p \alpha = e^{2\alpha}$. The solution can be expressed in terms of two arbitrary functions $f(z)$ and $g(\bar z)$
\be
  \label{eq:alp}
  e^{2\alpha} = -{ f'(z)  g'(\bar z)  \over [f(z)-g(\bar z)]^2 } \, .
\ee
Positivity of the area density requires $f'(z)g'(\bar z)<0$. We further assume $f(z)<g(\bar z)$.
The solutions of the scattering problem in (\ref{eq:linear}) can be written as
\be
  \label{eq:spinor1}
  \psi_{\alpha a}  = \le[g(\bar z)-f(z)\ri]^{-{1\ov 2}}\begin{pmatrix}
\quad \le[ c^1_a+ c^0_a f(z) \ri]  \zeta^{-{1\ov 2}}\le(-{g'(\bar z) \over f'(z)}\ri)^{1\ov 4}
\cr
-\le[c^1_a + c^0_a g(\bar z)  \ri] \zeta^{{1\ov 2}} \le(-{f'(z) \over g'(\bar z)}\ri)^{1\ov 4}
\end{pmatrix}  \, .
\ee
where $a \in \{0,1\}$ labels the two linearly independent solutions, $\alpha  \in \{0,1\}$ denotes the spinor index and $c^i_a$ are four arbitrary real constants.
The solution is valid in the vicinity of an AdS$_2$ patch. In Figure \ref{fig:wsheet} (right) the boundary of this region is indicated by a dashed line.

\subsection{Kink collision vertices}

\begin{figure}[h]
\begin{center}
\includegraphics[width=3.7cm]{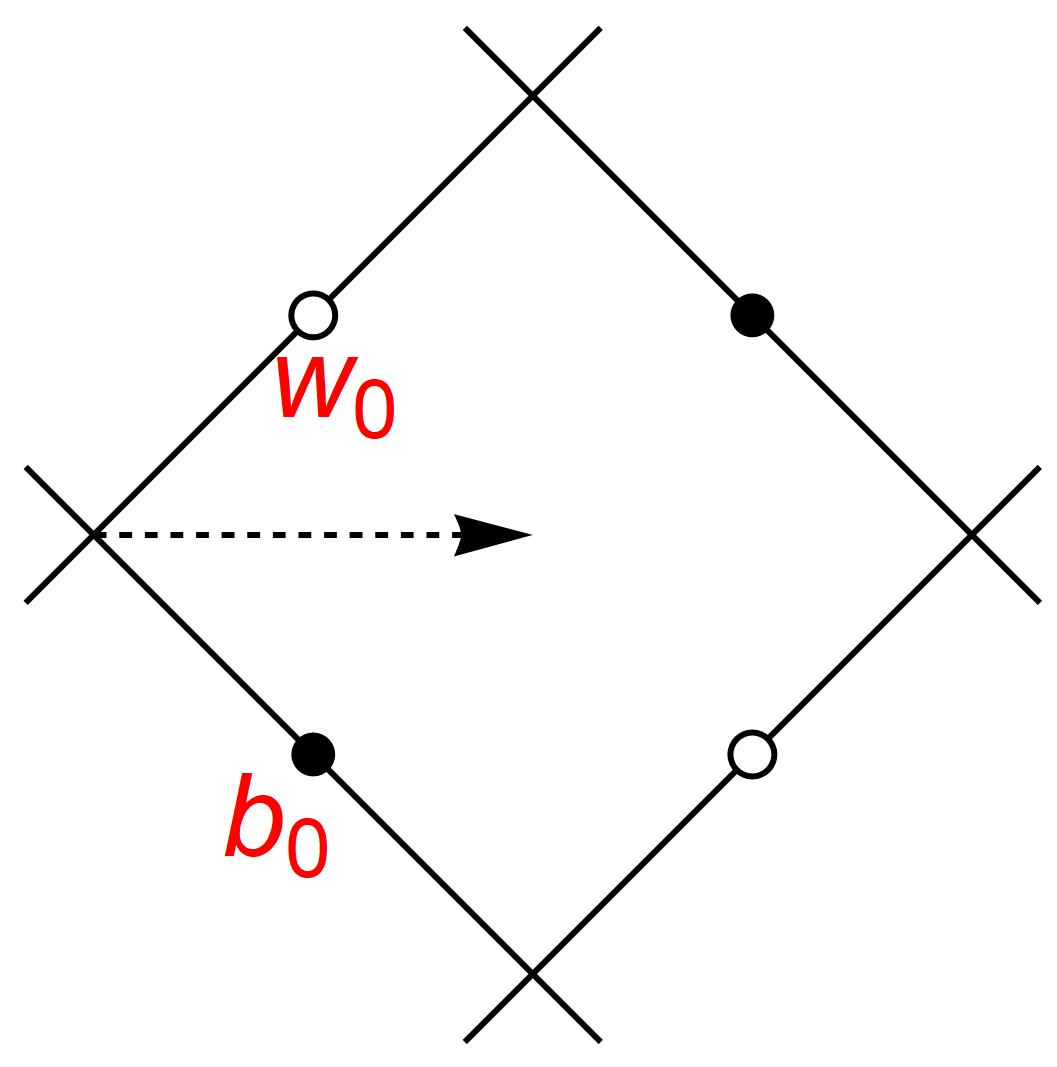} \qquad
\includegraphics[width=5.8cm]{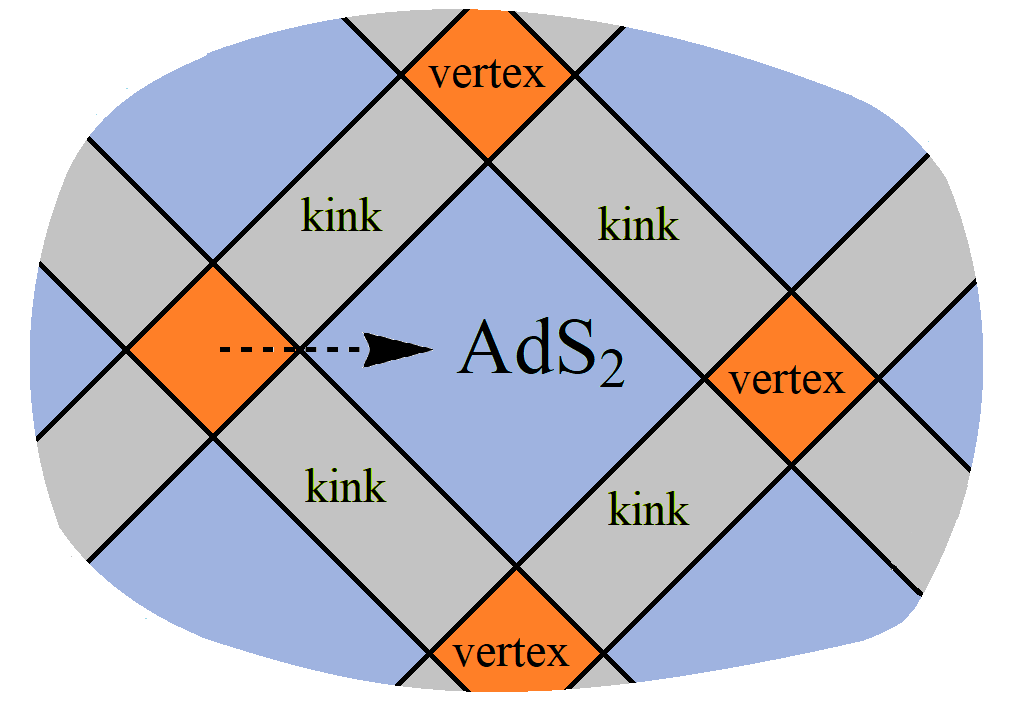} \qquad \includegraphics[width=3.2cm]{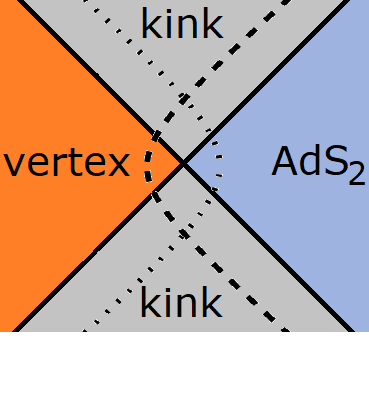}
\caption{\label{fig:wsheet} {\it Left:}  The arrow indicates the path for which the connection matrix can be computed in terms of the $b_0$ and $w_0$ celestial fields.
  {\it Middle:}  The same worldsheet patch after a conformal transformation. {\it Right:} The  $\psi_{\alpha a}$ and $ \tilde\psi_{\alpha a} $ solutions of the AdS$_2$ and vertex regions are valid to the right of the dashed line and to the left of the dotted line, respectively. 
}
\end{center}
\end{figure}

There is a symmetry between AdS$_2$ and vertex regions with the area and dual area densities exchanged.
Note that in a coordinate system where $u=v=\textrm{const.}$, a new solution to~eq.~(\ref{eq:linear}) can be obtained by the transformation that simultaneously exchanges the following quantities:
\be
  \nonumber
  \alpha \leftrightarrow \beta \, , \qquad z \leftrightarrow \bar z  \, , \qquad  \zeta \leftrightarrow \zeta^{-1} \, .
\ee
Hence, similarly to (\ref{eq:alp}), we can write the dual area density as
\be
  \label{eq:blp}
  e^{2\beta} = -{ \tilde f'(z)  \tilde g'(\bar z)  \over [\tilde f(z)-\tilde g(\bar z)]^2 } \, .
\ee
Let us parametrize the spinors solution in the vertex regions as
\be
  \label{eq:spinor2}
  \tilde\psi_{\alpha a}  =  [ \tilde f(z)-\tilde g(\bar z) ]^{-{1\ov 2}}
  \begin{pmatrix}
\le[\tilde c^1_a + \tilde c^0_a \tilde g(\bar z) \ri] \zeta^{1\ov 2} \le(-{\tilde f'(z) \over \tilde g'(\bar z)}\ri)^{1\ov 4}
\cr
-[\tilde c^1_a + \tilde c^0_a \tilde f(z) ] \zeta^{-{1\ov 2}} \le(-{\tilde g'(\bar z) \over \tilde f'(z)}\ri)^{1\ov 4}
\end{pmatrix}  \, ,
\ee
where tildes have been added to the various quantities to distinguish them from those appearing in (\ref{eq:spinor1}). The solution is valid in the vicinity of a vertex patch. In Figure \ref{fig:wsheet} (right) the boundary of this region is indicated by a dotted line.

\clearpage

\subsection{Local frame}

In the following, we will specify the $f, \tilde f, g, \tilde g$ functions, which amounts to choosing a local frame for the spinors. The global solution will then be determined by the $c^i_a$ and $\tilde c^j_b$ variables which can be thought of as piecewise constant functions on the union of AdS$_2$ and vertex patches, respectively. Note that the kink regions (gray areas in Figure \ref{fig:wsheet} (middle)) play no role in the construction.

The spin frame will be defined using the $\zeta=1$ spinors solutions (see Appendix A in \cite{Alday:2009yn}). These are physical in the sense that they determine the string embedding via (\ref{eq:physical}).  Equations (\ref{eq:bdef}) and  (\ref{eq:wdef}) define the physical celestial variables which can be used to fix $f, \tilde f, g, \tilde g$. Generically $b= b(z, \bar z)$ and $w=w(z, \bar z)$. This is simplified in the vertex and AdS$_2$ regions of a segmented string: the celestial variables only depend on either $z$ or $\bar z$ \cite{Vegh:2019any}. Thus, we can set
\be
  \label{eq:setfg}
   f(z) = w(z) \, , \quad g(\bar z) = b(\bar z) \, ,
   \quad  c_a^i = \delta_a^i 
\ee
on AdS$_2$ patches and
\be
  \label{eq:setfgt}
  \tilde f(z) = b(z) \, , \quad \tilde g(\bar z) = w(\bar z)   \, ,
   \quad \tilde c_a^i = \delta_a^i 
\ee
on vertex patches. Note that with this choice, the solutions are properly normalized according to (\ref{eq:normalize}).
By plugging these functions into (\ref{eq:alp}) and (\ref{eq:blp}) one can see that they indeed give the correct  area and dual area densities \cite{Vegh:2019any}
\be
  \nonumber
  e^{2\alpha} = -{ \bar\p b  \p w  \over (b-w)^2 } \, , \qquad
  e^{2\beta} = -{ \p b  \bar\p w  \over (b-w)^2 } \, .
\ee

\begin{figure}[h]
\begin{center}
\includegraphics[width=4.5cm]{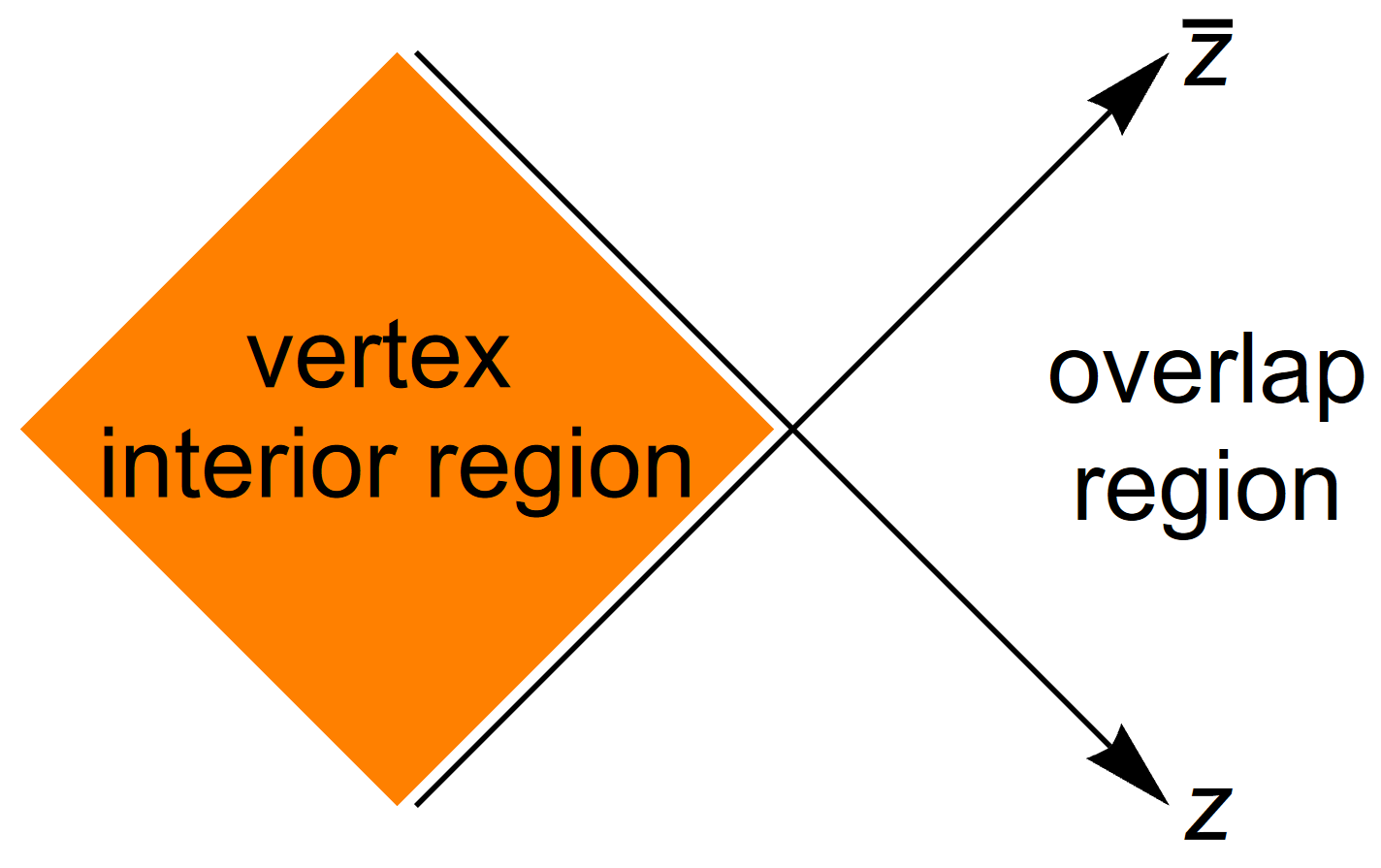} \qquad
\includegraphics[width=4.5cm]{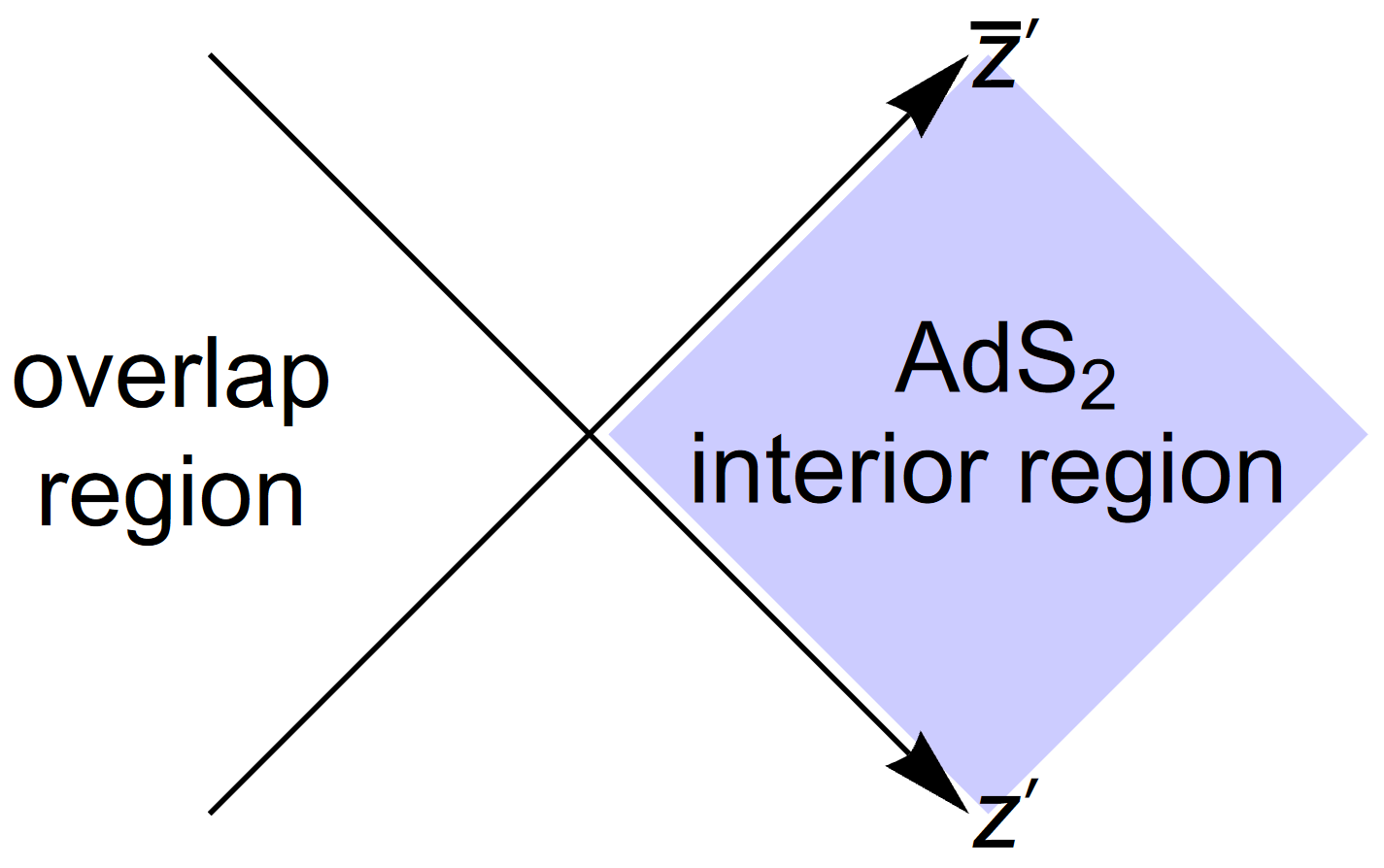} \
\caption{\label{fig:coosys} Coordinate systems for the matched asymptotic expansion.
}
\end{center}
\end{figure}

\subsection{Matching the solutions}

The spinor solutions (\ref{eq:spinor1}) and (\ref{eq:spinor2}) are only valid strictly inside  AdS$_2$ and vertex patches, respectively.
In order to match them, we need to define them in slightly larger regions as seen in Figure \ref{fig:wsheet} (right). This is still possible within the framework of the Liouville equation. Let us consider, for instance, a vertex patch. This patch, along with a small surrounding area on the worldsheet (to the left of the dotted line in the Figure) is mapped into the vicinity of a spacetime point. After zooming in, the target space curvature is scaled out and what is left is a piecewise linear string in Minkowski space satisfying the Liouville equation (see section 2 in \cite{Jevicki:2009uz}). Similarly, the Liouville equation is capable of describing AdS$_2$ patches with their infinitesimally small neighborhoods (everything to the right of the dashed line in Figure \ref{fig:wsheet} (right)). In the intersection of the two regions, both non-linear terms in the sinh-Gordon equation vanish and thus we have $\p\bar\p \alpha = 0$ (and similarly  $\p\bar\p \beta = 0$).

We would like to make sure that the solutions are correctly glued and that the spin frame is globally well-defined. This requires that in the overlap region 
\be
  \nonumber
   \psi_{\alpha, 0}({\zeta=1}) =  \tilde\psi_{\alpha, 0}({\zeta=1}) \,  \quad \textrm{and} \quad
   \psi_{\alpha, 1}({\zeta=1}) =  \tilde\psi_{\alpha, 1}({\zeta=1}) \, .
\ee
Let us consider two separate coordinate systems $(z, \bar z)$ and $(z', \bar z')$  for the vertex and AdS$_2$ patches, respectively. In the coordinate system for the vertex and AdS$_2$ patches the overlapping regions are assumed to lie in the quadrants $z, \bar z > 0$ and $z', \bar z' < 0$, respectively; see Figure \ref{fig:coosys}. The opposite quadrants  ($z, \bar z < 0$ and $z', \bar z' > 0$) are the interior regions of the patches: these are the orange and blue regions in Figure \ref{fig:wsheet} (right). We choose the coordinates so that in the interior regions  $b$ and $w$ are fractional linear functions (of the appropriate single variable), and the $f, \tilde f, g, \tilde g$ functions are set equal to them according to (\ref{eq:setfg}) and (\ref{eq:setfgt}).

In the overlapping regions, $f, \tilde f, g, \tilde g$ should asymptote to constants
\be
  \nonumber
  \lim_{z' \to -\infty} f(z') = \lim_{\bar z \to +\infty} \tilde g(\bar z) = w_0 \, ,  \qquad  \lim_{z \to +\infty} \tilde f(z) = \lim_{\bar z' \to -\infty}   g(\bar z') =   b_0 \, .
\ee
As a simple example let us consider the functions
\be
  \nonumber
  f(z') =  w_0 - \varepsilon\log(1+e^{z' / \varepsilon}) \,   \qquad  \textrm{and} \qquad
  g(\bar z') = b_0 + \varepsilon\log(1+e^{\bar z' / \varepsilon})
\ee
which interpolate between  $w_0$ and $w_0-z'$ and between $b_0$ and $b_0+\bar z'$, respectively. Here $\varepsilon$ is a small parameter.
Similarly, consider
\be
  \nonumber
  \tilde f(z) =  b_0 + \varepsilon\log(1+e^{-z / \varepsilon}) \,   \qquad  \textrm{and} \qquad
  \tilde g(\bar z) = w_0 - \varepsilon\log(1+e^{-\bar z / \varepsilon}) \, ,
\ee
which interpolate between  $b_0-z$ and $b_0$ and between $w_0+\bar z$ and $w_0$, respectively.
The corresponding spinors (with $\zeta=1$) in the overlapping region are
\be
  \nonumber
  \psi_{\alpha, 0}  \approx (b_0 - w_0)^{-{1\ov 2}}\begin{pmatrix}
w_0 \, e^{{(\bar z' - z')/( 4 \varepsilon)}}
\cr
-b_0 \, e^{{( z' - \bar z')/( 4 \varepsilon)}}
\end{pmatrix}  \, , \qquad
\psi_{\alpha, 1}  \approx (b_0 - w_0)^{-{1\ov 2}}\begin{pmatrix}
  e^{{(\bar z' - z')/( 4 \varepsilon)}}
\cr
 - e^{{( z' - \bar z')/( 4 \varepsilon)}}
\end{pmatrix}  \, ,
\ee
\be
  \nonumber
  \tilde\psi_{\alpha, 0}  \approx (b_0 - w_0)^{-{1\ov 2}}\begin{pmatrix}
w_0 \, e^{{(\bar z - z)/( 4 \varepsilon)}}
\cr
-b_0 \, e^{{( z - \bar z)/( 4 \varepsilon)}}
\end{pmatrix}  \, , \qquad
\tilde\psi_{\alpha, 1}  \approx (b_0 - w_0)^{-{1\ov 2}}\begin{pmatrix}
  e^{{(\bar z - z)/( 4 \varepsilon)}}
\cr
 - e^{{( z - \bar z)/( 4 \varepsilon)}}
\end{pmatrix}  \, .
\ee
Identifying $z-\bar z = z' - \bar z'$ we see that $\psi_{\alpha,0}  =\tilde\psi_{\alpha, 0}$ and $\psi_{\alpha,1}  =\tilde\psi_{\alpha, 1}$. Note that they are not constants, but their dependence on the coordinates are the same. The example can   easily be generalized to interpolating functions connecting fractional linear functions to $b_0$ and $w_0$.

For $\zeta \ne 1$ the Lax connection matrix will generally be non-trivial. We can compute it by equating   $\psi$ and $\tilde\psi$ in the overlapping region which defines the discrete Lax matrix $\Omega$
\be
  \label{eq:defomega}
   \psi_{\alpha a} = \Omega_{ab}(b_0, w_0, \zeta) \tilde\psi_{\alpha b}
    \, , \qquad \textrm{or} \qquad
    \psi = \tilde\psi \Omega^T\, ,
\ee
where the spinors can be written as
\be
  \nonumber
   \psi  = (b_0-w_0)^{-\half}
   \begin{pmatrix}
 \le(-{g'(\bar z') \over f'(z')}\ri)^{1\ov 4} & 0 \cr
 0  &  \le(-{f'(z') \over g'(\bar z')}\ri)^{1\ov 4}
\end{pmatrix}
 \begin{pmatrix}
 \zeta^{-\half} & 0 \cr
 0  &  \zeta^{\half}
\end{pmatrix}
 \begin{pmatrix}
w_0   & 1  \cr
-b_0  & -1
\end{pmatrix} \, ,
\ee
\be
   \nonumber
  \tilde\psi  =   (b_0-w_0)^{-\half}
   \begin{pmatrix}
 \le(-{\tilde f'(z) \over \tilde g'(\bar z)}\ri)^{1\ov 4} & 0 \cr
 0  &  \le(-{\tilde g'(\bar z) \over \tilde f'(z)}\ri)^{1\ov 4}
\end{pmatrix}
 \begin{pmatrix}
 \zeta^{\half} & 0 \cr
 0  &  \zeta^{-\half}
\end{pmatrix}
 \begin{pmatrix}
w_0   & 1  \cr
-b_0  & -1
\end{pmatrix} \, .
\ee
As we have seen, in the overlap regions we have ${g'(\bar z') \over f'(z')} = {\tilde f'(z) \over \tilde g'(\bar z)}$. Thus, (\ref{eq:defomega}) simplifies to
\be
  \nonumber
  \begin{pmatrix}
 \zeta^{-\half} & 0 \cr
 0  &  \zeta^{\half}
\end{pmatrix}
 \begin{pmatrix}
w_0   & 1  \cr
-b_0  & -1
\end{pmatrix} =
 \begin{pmatrix}
 \zeta^{\half} & 0 \cr
 0  &  \zeta^{-\half}
\end{pmatrix}
 \begin{pmatrix}
w_0   & 1  \cr
-b_0  & -1
\end{pmatrix} \Omega^T \, .
\ee
This is solved by
\be
  \nonumber
\Omega_{b_0,w_0} = {1 \over (b_0-w_0) \zeta}
\begin{pmatrix}
{b_0 \zeta^2 - w_0 } & {b_0 w_0(1-\zeta^2)  }
\cr
 { \zeta^2-1  } & { b_0 -w_0\zeta^2  }
\end{pmatrix} \, ,
\ee
where we have adopted the notation that the celestial fields are written as a subscript (not to be confused with the spacetime spinor indices in (\ref{eq:defomega})).

\clearpage

\section{The discrete Lax matrix}

In the previous section we have computed the connection matrix for a path that started on an AdS$_2$ patch and ended on an adjacent vertex (see dashed line in Figure \ref{fig:wsheet} (left)). The result depends on the two nearest celestial variables
\be
  \label{eq:mono}
\Omega_{b,w}(\zeta) =
{1 \over (b-w) \zeta}
\begin{pmatrix}
{b \zeta^2 - w } & {b w(1-\zeta^2)  }
\cr
 { \zeta^2-1  } & { b -w\zeta^2  }
\end{pmatrix}
\ee
The Lax matrix is in $SL(2, \RR)$
\be
  \nonumber
  \det \Omega_{b,w} = 1 \, ,
\ee
and without the spectral parameter it is equal to the $2\times 2$ identity matrix by construction
\be
   \nonumber
\Omega_{b,w}(\zeta=1) = \mathbb{1} \, .
\ee

\noindent
Although in the example in Figure \ref{fig:wsheet} the path goes `east',  (\ref{eq:mono}) gives the connection matrix for the other three directions as well. Note that the path is always pointing from the vertex to the inside of the AdS$_2$ patch. The Lax matrix of the reverse path is given by
\be
  \nonumber
  \Omega_{b,w}^{-1}(\zeta) =   \Omega_{w,b}(\zeta)  =   \Omega_{b,w}(\zeta^{-1})\, .
\ee
Simultaneous M\" obius transformations of the celestial variables
\be
  \nonumber
  b \to b' = {a_{00} b + a_{01} \over a_{10} b + a_{11} } \, , \qquad  w \to w' = {a_{00} w + a_{01}  \over a_{10} w + a_{11} }
\ee
are isometry transformations of AdS$_3$. The connection matrix transforms covariantly
\be
  \nonumber
  \Omega_{b,w} \to  U \Omega_{b',w'} U^{-1} \, \qquad \textrm{with} \qquad U =  \begin{pmatrix}
a_{00} & a_{01}
\cr
a_{10} & a_{11}
\end{pmatrix}   \, .
\ee

\clearpage

\begin{figure}[h]
\begin{center}
\includegraphics[width=4cm]{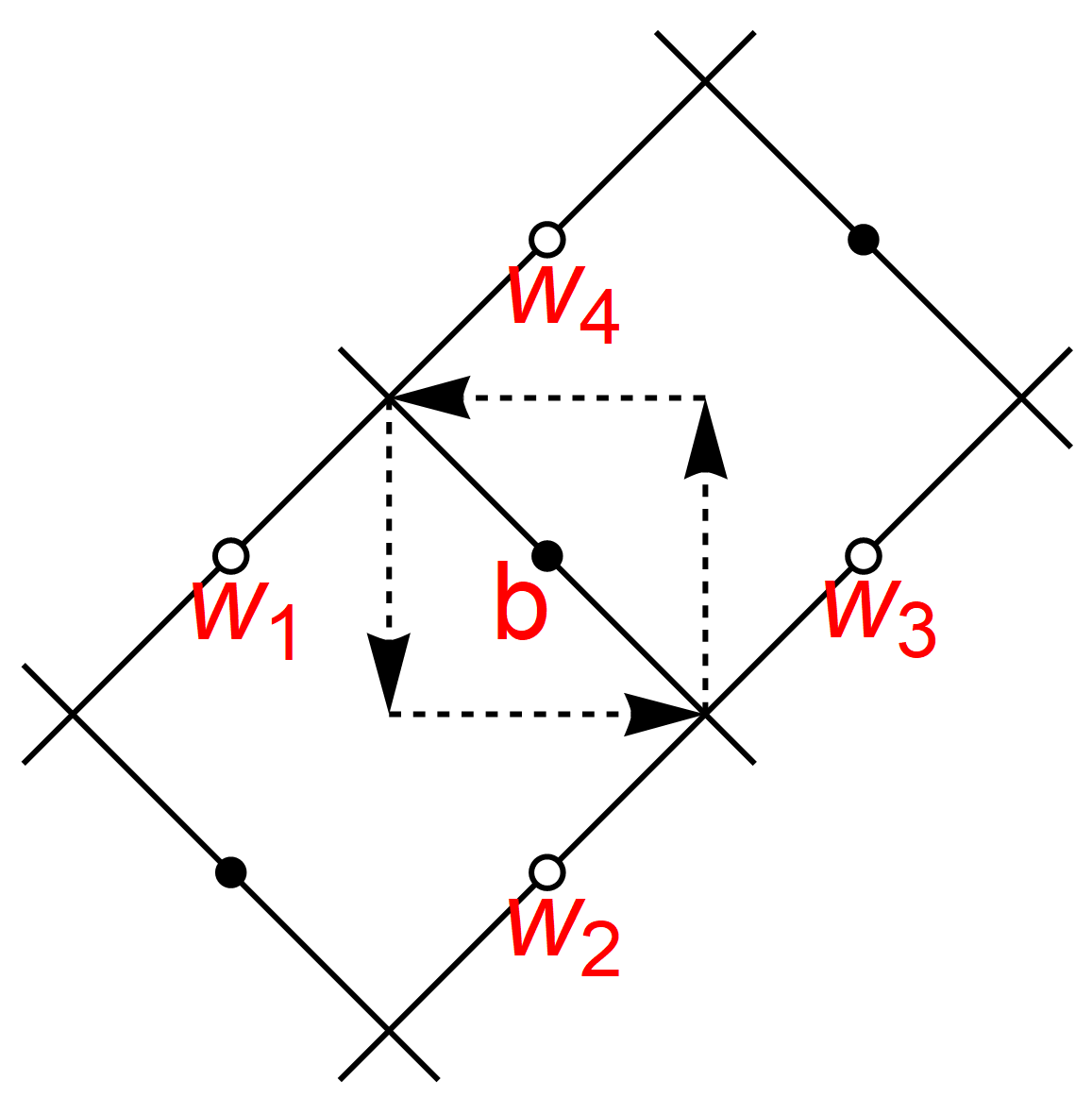} \qquad\qquad
\includegraphics[width=4cm]{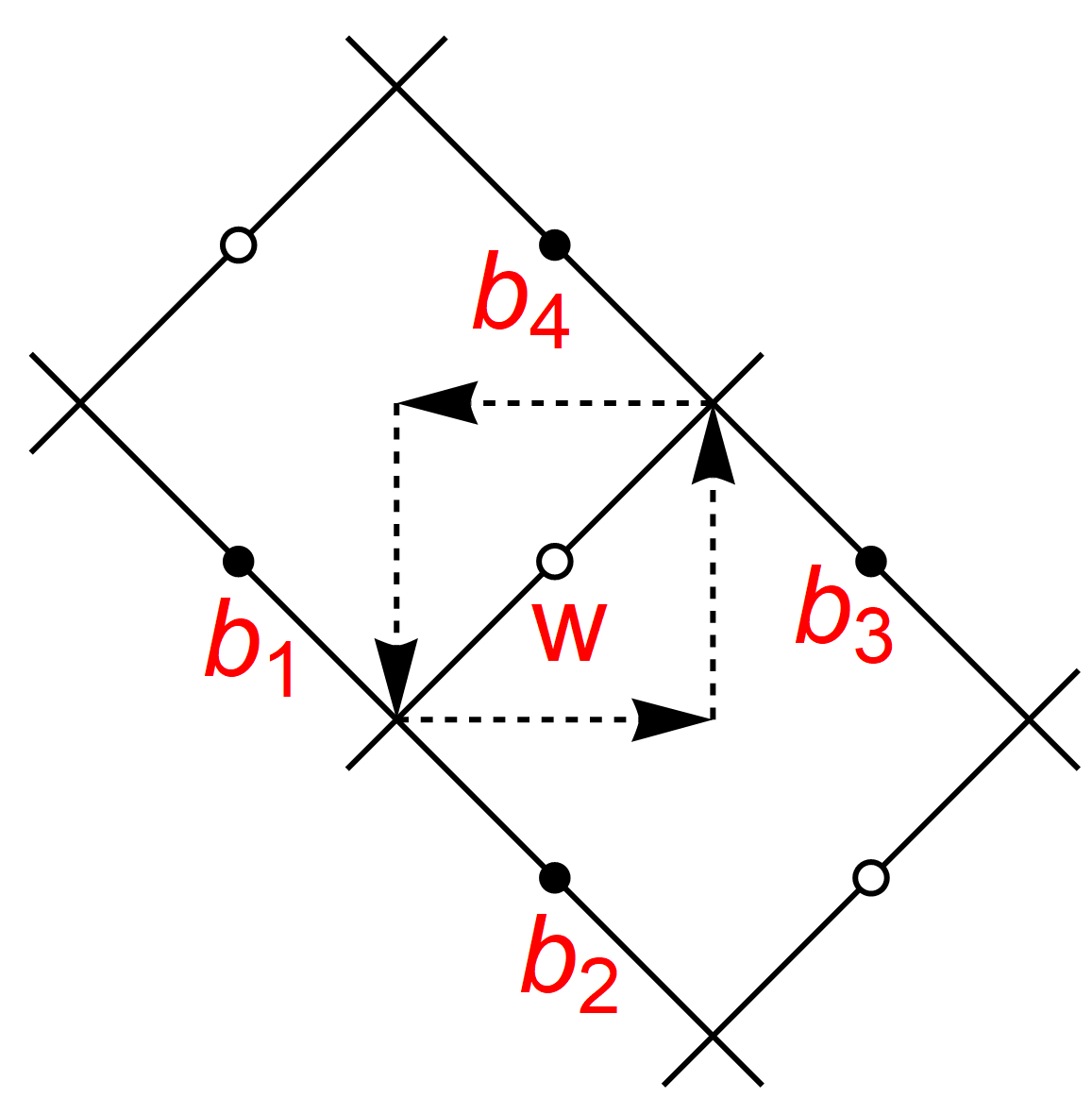}
\caption{\label{fig:flat} Monodromy around trivial loops. The (\ref{eq:deqn}) equation of motion relates $b$ and the four $w_i$ variables (and similarly $w$ and the four $b_i$) ensuring the flatness of the connection.
}
\end{center}
\end{figure}

\subsection{Flatness}

The connection is flat and thus it is trivial around contractible loops. We will first show this for the smallest  loops which consist of four elementary paths as in Figure \ref{fig:flat}.  Assuming the $b$, $w_1, w_2, w_3, w_4$ celestial variables satisfy the equation of motion in (\ref{eq:deqn}), i.e.
\be
  \nonumber
  {1\ov b - w_2}+   {1\ov b - w_4} =
  {1\ov b - w_1}+   {1\ov b - w_3} \, ,
\ee
we get
\be
  \nonumber
  \Omega_{b,w_4}^{-1}\Omega_{b,w_3}\Omega_{b,w_2}^{-1}\Omega_{b,w_1} = \mathbb{1} \,
\ee
for any $\zeta$. The monodromy around a white dot also gives a similar trivial result,
\be
  \nonumber
  \Omega_{b_4,w}\Omega_{b_3,w}^{-1}\Omega_{b_2,w}\Omega_{b_1,w}^{-1} = \mathbb{1} \, ,
\ee
if the equation of motion is used again.
Larger contractible loops can be built from the two small loops. Hence the monodromy for all of these will be trivial.

\clearpage

\begin{figure}[h]
\begin{center}
\includegraphics[width=3.0cm]{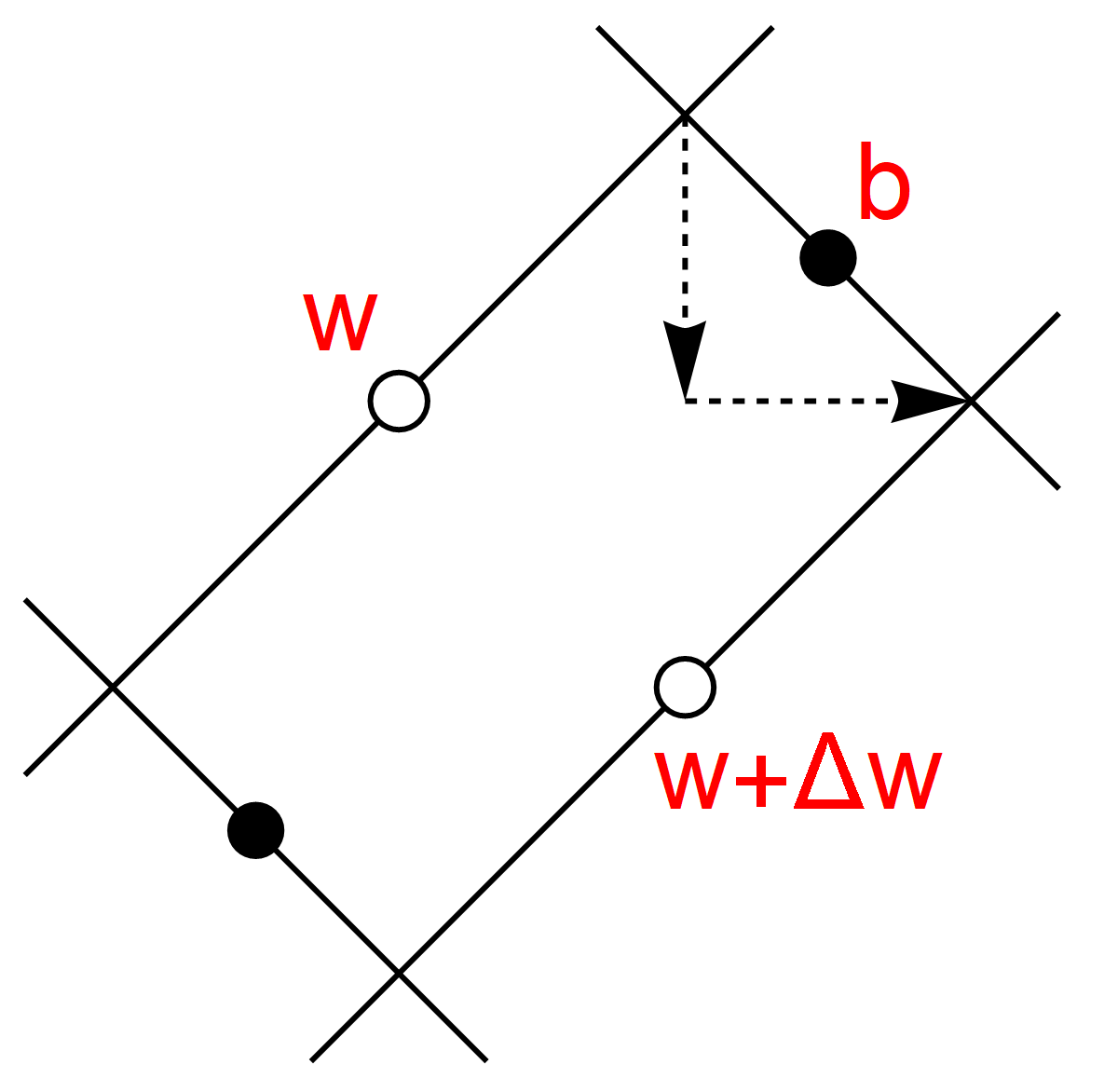} \qquad
\includegraphics[width=3.0cm]{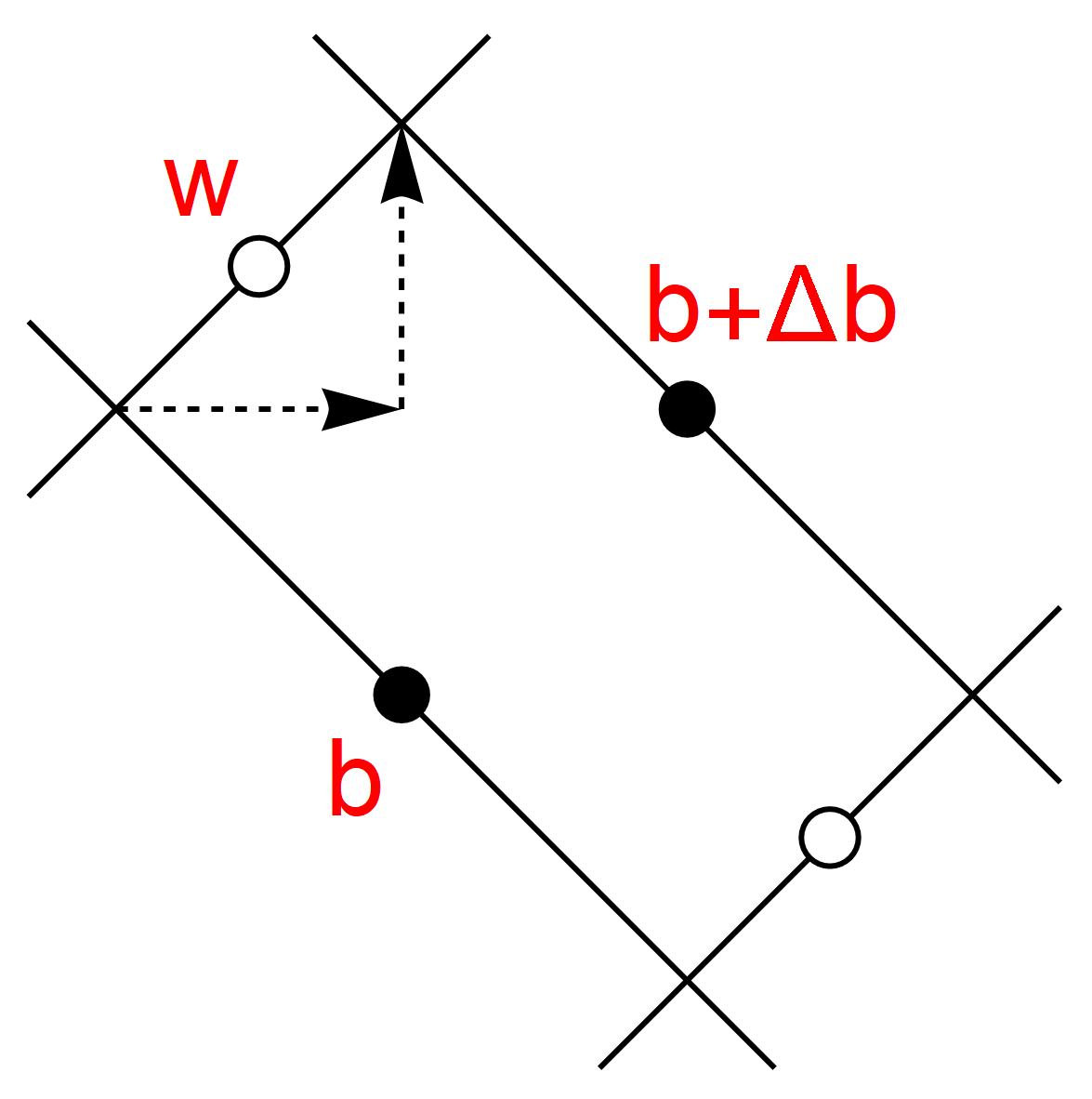} \qquad\qquad
\includegraphics[width=3.0cm]{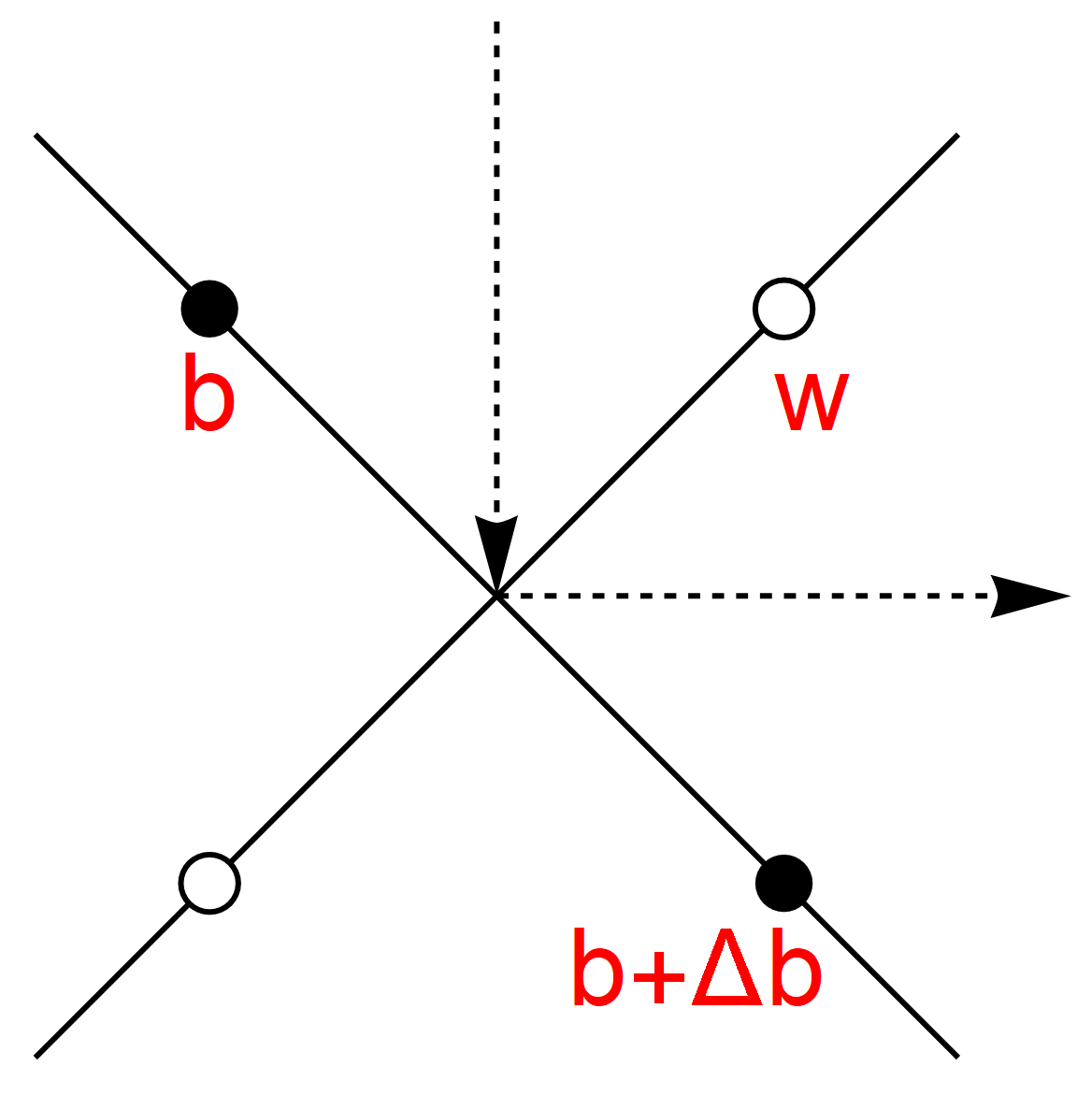} \qquad
\includegraphics[width=3.0cm]{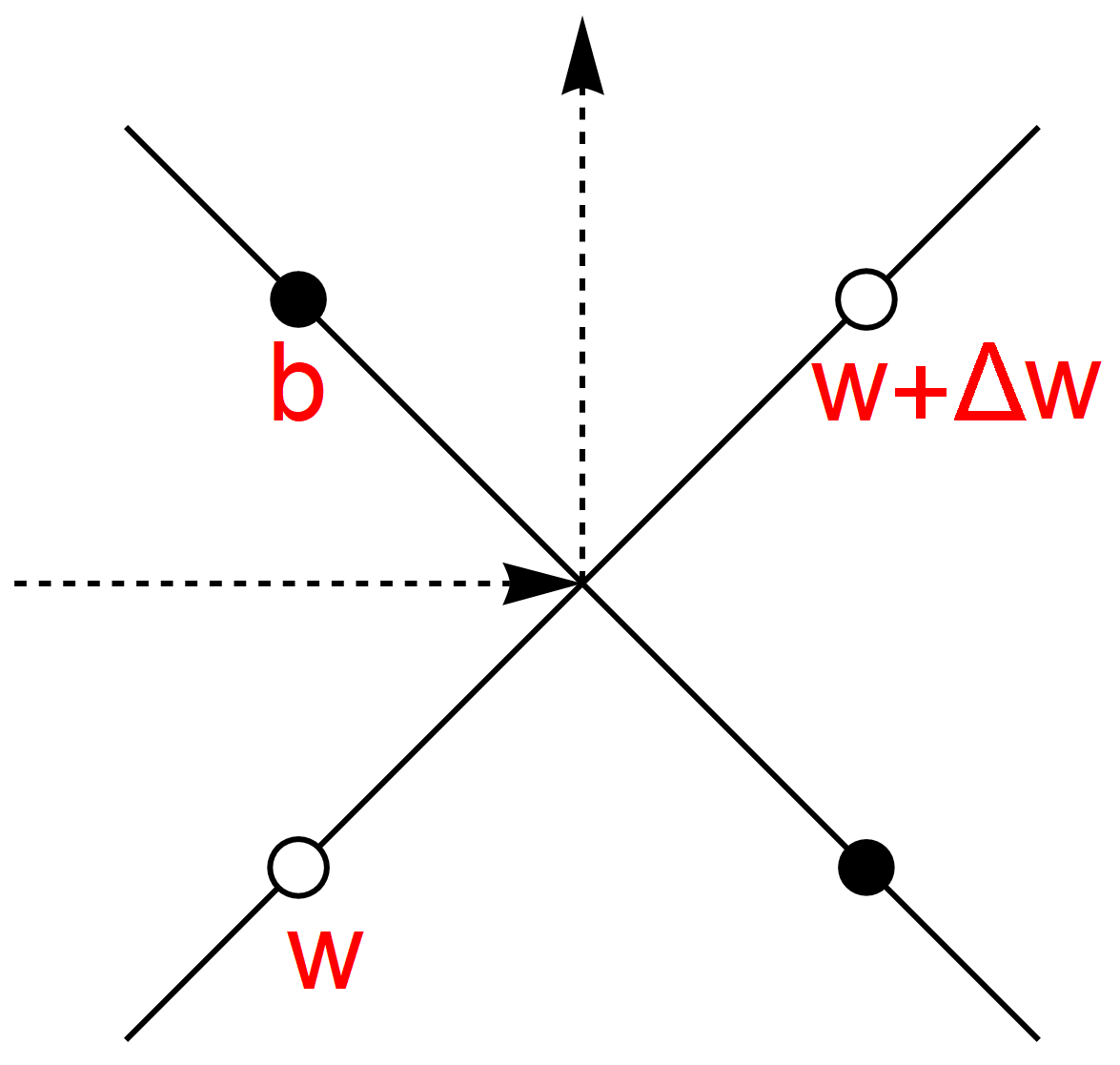}
\caption{\label{fig:limit} Taking the continuum limit of the discrete Lax  matrix.
}
\end{center}
\end{figure}

\subsection{Continuum limit}

Taking the continuum limit of $\Omega_{b, w}$ is not immediately possible, because the matrix depends on two variables $b$ and $w$ that are only equal when the string touches the AdS boundary. One can, however, take the product of two Lax matrices as shown in the first two diagrams in Figure \ref{fig:limit}. For small $\Delta w \approx \p w  \, \Delta z$ we have
\be
  \label{eq:cont1}
  \Omega^{-1}_{b, w+\Delta w} \Omega_{b, w} \approx \mathbb{1}  +{ j_{  z}\over 1-x}  \Delta z \, , \qquad
  j_{  z}=
 {2\over (b-w)^2 }   \begin{pmatrix}
-b \p w & b^2 \p w
\cr
-\p w & b \p w
\end{pmatrix}   \, ,
\ee
and similarly in the $\bar z$ direction (with $\Delta b \approx \bar\p b  \, \Delta \bar z$)
\be
  \label{eq:cont2}
  \Omega^{-1}_{b+\Delta b, w} \Omega_{b, w} \approx \mathbb{1} +{ j_{\bar z}\over 1+x}  \Delta \bar z \, , \qquad
  j_{\bar z}=
 {2\over (b-w)^2}   \begin{pmatrix}
-w \bar\p b & w^2 \bar\p b
\cr
-\bar\p b & w \bar\p b
\end{pmatrix}   \, .
\ee
Here we have introduced a new spectral parameter $x$, which usually appears in the definition of the Lax connection of the principal chiral model. It is related to $\zeta$ by
\be
  \label{eq:xdef}
  \zeta = \sqrt{x-1 \ov x+1} \, .
\ee

Another pair of matrices can be obtained from the paths that start and end on adjacent AdS$_2$ patches; see the diagrams on the RHS of Figure \ref{fig:limit}. From these paths we get
\be
  \nonumber
  \Omega_{b+\Delta b,w} \Omega^{-1}_{b, w} \approx \mathbb{1}  +{\tilde\jmath_{  z}\over 1-x}  \Delta z \, , \qquad
 \tilde\jmath_{  z}=
 {2\over (b-w)^2 }   \begin{pmatrix}
-w \p b & w^2 \p b
\cr
-\p b & w \p b
\end{pmatrix}   \, ,
\ee
\be
  \nonumber
  \Omega_{b, w+\Delta w} \Omega^{-1}_{b, w} \approx \mathbb{1} +{\tilde\jmath_{\bar z}\over 1+x}  \Delta \bar z \, , \qquad
  \tilde\jmath_{\bar z}=
 {2\over (b-w)^2}   \begin{pmatrix}
-b \bar\p w & b^2 \bar\p w
\cr
-\bar\p w & b \bar\p w
\end{pmatrix}   \, .
\ee
which are the same as (\ref{eq:cont1}), (\ref{eq:cont2}) with $b$ and $w$ exchanged.

Spacetime isometries act as M\" obius transformations on the celestial variables. The two classically equivalent actions in (\ref{eq:bwaction}) are invariant under such transformations and $j$ and $\tilde\jmath$ are the associated $SL(2)$ currents. They satisfy the conservation laws and flatness conditions
\be
 \nonumber
\partial j_{\bar z} + \bar{\partial}j_z  = 0   \, , \qquad
\partial \tilde\jmath_{\bar z} + \bar{\partial}\tilde\jmath_z  = 0 \, ,
\ee
\be
\nonumber
\partial j_{\bar z} - \bar{\partial}j_z -[j_z, j_{\bar{z}}] = 0 \, , \qquad
\partial \tilde\jmath_{\bar z} - \bar{\partial}\tilde\jmath_z -[\tilde\jmath_z, \tilde\jmath_{\bar{z}}] = 0 \, .
\ee
The currents can be written in the usual form $j = -dY Y^{-1}$ where $Y_{a\dot a}$ is the embedding matrix in (\ref{eq:yeq}). For convenience, we repeat the discussion from Appendix A in \cite{Alday:2009yn}.
If we define the matrices
\be
  \nonumber
M_1^{\alpha \dot \beta}=\begin{pmatrix}
1 & 0 \cr
0  & 1
\end{pmatrix} \, , \qquad
M_2^{\alpha \dot \beta}=\begin{pmatrix}
0 & 1 \cr
0  & 0
\end{pmatrix} \, , \qquad
M_3^{\alpha \dot \beta}=\begin{pmatrix}
0 & 0 \cr
1  & 0
\end{pmatrix} \, , \qquad
\ee
then we have from (\ref{eq:wmat})
\bea
  \nonumber
 Y_{a \dot a }  &=&
\psi^L_{\alpha a} M_1^{\alpha \dot \beta} \psi^R_{\dot \beta  \dot{a}}  = \le[(\psi^L)^t M_1 \psi^R \ri]_{a \dot a } \, ,
\\
\nonumber
 Y^{-1}  &=&
 (\psi^R)^{-1} M_1^{-1} \le[(\psi^L)^t \ri]^{-1} \, ,
\\
\nonumber
 \p Y  &=&
 2 e^{\alpha}(\psi^L)^t  M_3 \psi^R  \, ,
\\
\nonumber
 \bar\p Y  &=&
 2 e^{\alpha}(\psi^L)^t  M_2 \psi^R  \, .
\eea
Using these expressions, the left current can be written as
\bea
\nonumber
  {(j_z)_a}^b=  - \p Y Y^{-1} = - 2 e^{\alpha}(\psi^L)^t  M_3 M_1^{-1} \le[(\psi^L)^t \ri]^{-1} \, .
\eea
If we now plug in
\be
  \nonumber
   \psi^L = (b-w)^{-\half}\begin{pmatrix}
w \le( -{ \bar\p b \ov  \p w} \ri)^{{1\ov 4}} & \le( -{ \bar\p b \ov  \p w} \ri)^{{1\ov 4}} \cr
-b\le( -{  \p w \ov \bar\p b } \ri)^{{1\ov 4}}  & -\le( -{  \p w \ov \bar\p b } \ri)^{{1\ov 4}}
\end{pmatrix} \, , \qquad e^\alpha = {\sqrt{ - \bar\p b\p w} \ov b-w} \, ,
\ee
then we obtain precisely  (\ref{eq:cont1}).
Similarly we have
\bea
\nonumber
  {(j_{\bar z})_a}^b=  - \bar\p Y Y^{-1} = - 2 e^{\alpha}(\psi^L)^t  M_2 M_1^{-1} \le[(\psi^L)^t \ri]^{-1} \, ,
\eea
which gives the matrix in (\ref{eq:cont2}).

\clearpage

\begin{figure}[h]
\begin{center}
\includegraphics[width=4cm]{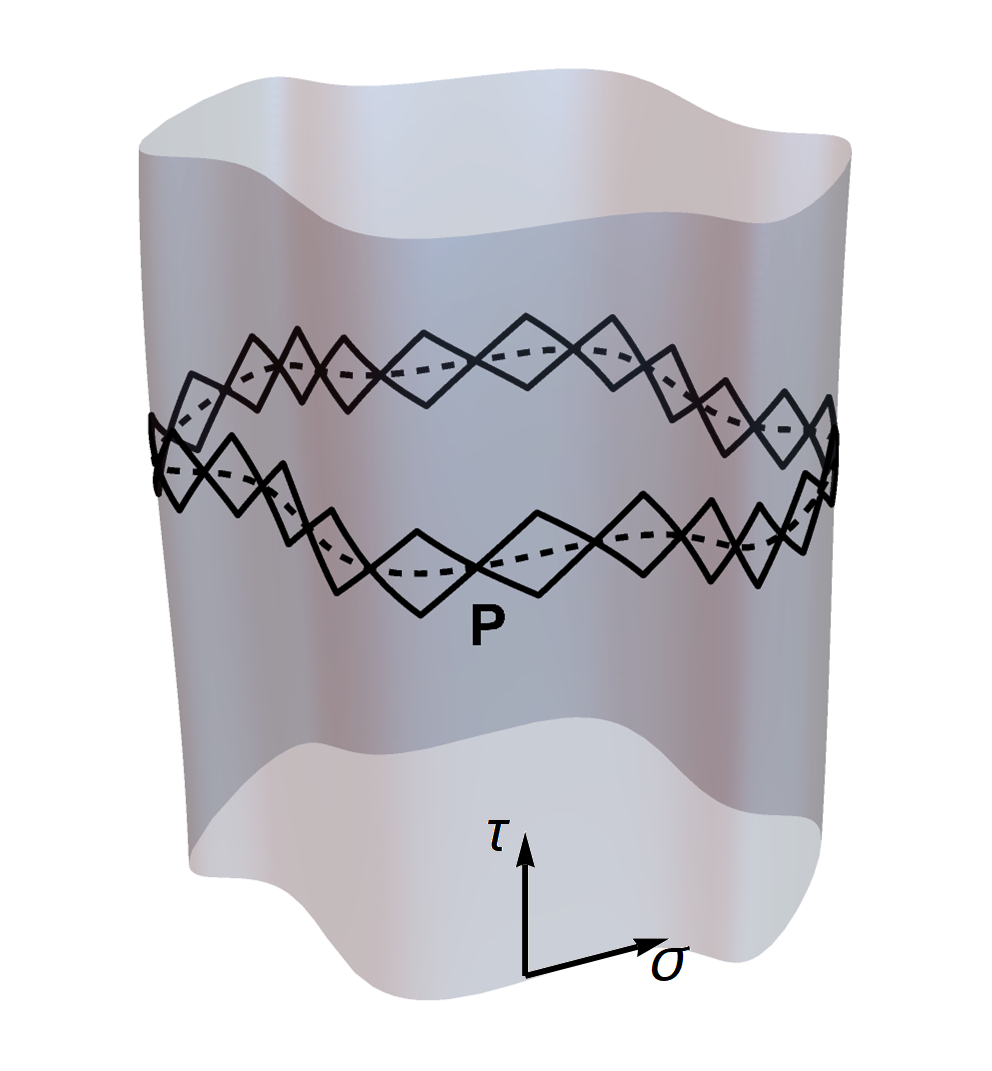} \qquad
\includegraphics[width=11cm]{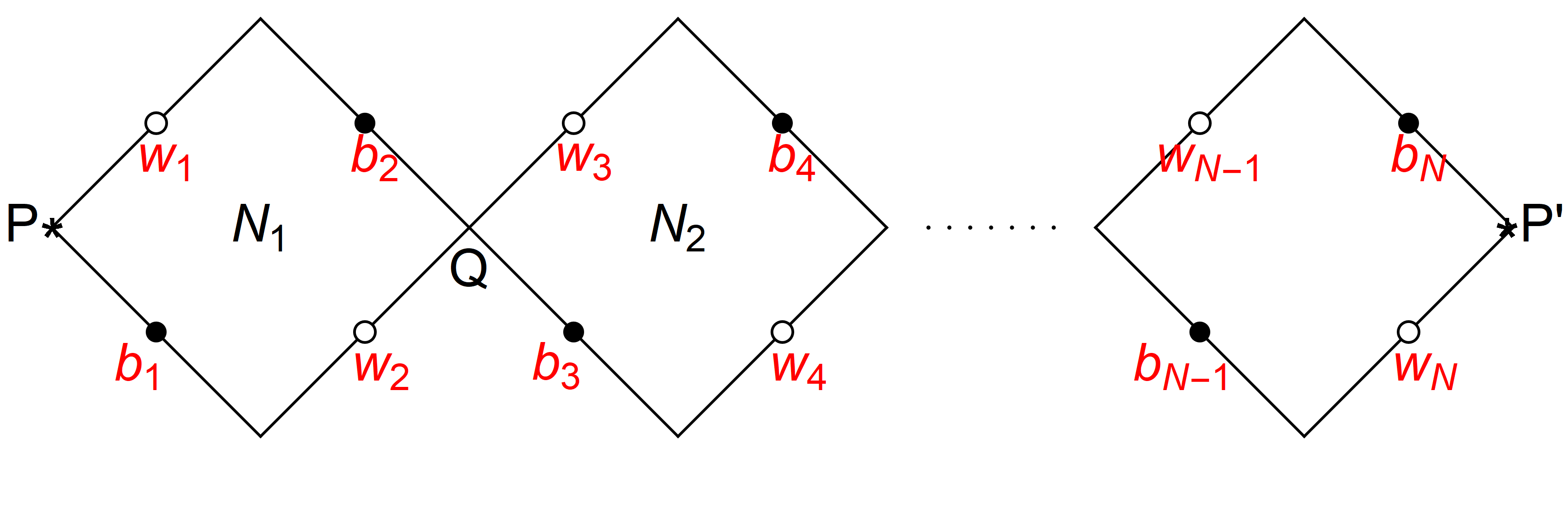}
\caption{\label{fig:spectral}{\it Left:} Schematic drawing of the worldsheet of a closed string with $N$ segments. Kink worldlines are indicated near a timeslice (dashed loop).  {\it Right:} Computing the monodromy around a closed loop. $P, Q, P' \in \RR^{2,2}$ denote kink collision vertices (we assume $P = P'$). $N_1$ and $N_2$ are the normal vectors of the first two AdS$_2$ patches.
}
\end{center}
\end{figure}

\section{The monodromy}

The spectral curve is obtained by computing the monodromy of the Lax connection around the non-trivial loop on the string worldsheet. This path and nearby kink worldlines are shown in Figure \ref{fig:spectral}. Let us assume that on a generic time-slice there are an even number of string segments with ${N/ 2}$ left-moving and ${N/ 2}$ right-moving kinks\footnote{The diagram assumes that the number of left- and right-moving kinks are the same. If this is not the case, then some of the kinks need to be removed by constraining the corresponding celestial variables (e.g. if we set $b_2 = b_3$, then the corresponding right-moving kink with $w_{2,3}$ variables will vanish).}. Then, the monodromy is a product of Lax matrices from $P$ to $P'$ given by
\be
  \label{eq:monodromy}
\Omega(\zeta) = \Omega_{b_N, w_N}^{-1} \Omega_{b_{N-1}, w_{N-1}} \cdots \Omega_{b_2, w_2}^{-1} \Omega_{b_1, w_1} \, .
\ee
This formula gives an explicit expression for the monodromy in terms of the celestial variables associated with the kink velocities. Note that the monodromy depends on the starting point $P$ on the worldsheet. If this point is changed to $\tilde P$, then $\Omega$ changes by conjugation,
\be
  \nonumber
  \Omega_P \to \Omega_{\tilde P} = U \Omega_P U^{-1}
\ee
where $U$ is the Lax matrix connecting $P$ and $\tilde P$.

\clearpage
\subsection{Closing the string}

For generic values of the $b_1, w_1, \ldots, b_N, w_N$ celestial variables the string does not close. In the following we will compute $P'$ from $P$ (both are considered to be points in $\RR^{2,2}$) by a series of transformations and constrain the celestial variables by requiring $P=P'$. First we compute the nearest normal vector
\be
   \nonumber
 N_1 =  \mathcal{R}_{b_1,w_1} P
\ee
where the `reflection matrix' is given by \cite{Vegh:2019any}
\be
\label{eq:reflmat}
\mathcal{R}_{b,w} =
{1\ov b-w}
\begin{pmatrix}
0 & bw+1 & bw-1 & -b-w
\cr
-1-bw & 0 & b+w & bw-1
\cr
-1+bw & b+w & 0 & -1-bw
\cr
-b-w & bw-1 & bw+1 & 0
\end{pmatrix}   \, .
\ee
From $N_1$ we can compute the next vertex $Q$ via
\be
   \nonumber
 Q =  \mathcal{R}_{b_2,w_2} N_1 \, .
\ee
From $Q$ we compute $N_2$ and so on until we get to $P'$. The final result contains the product of   matrices
\be
  \label{eq:product}
   \mathcal{R} =   \mathcal{R}_{b_N,w_N} \cdots \mathcal{R}_{b_2,w_2} \mathcal{R}_{b_1,w_1} \, .
\ee
If $\mathcal{R}  = \textrm{Id}_{4\times 4}$ then we have $P=P'$.
It turns out that this product can be repackaged into a product of $\Omega_{b,w}$  matrices as follows.
Let us consider the $n^\textrm{th}$ matrix in the product. It can be decomposed as
\be
   \nonumber
  \mathcal{R}_{b_n,w_n} =  c^\mu_n \Sigma_\mu \, ,
\ee
where summation over $\mu \in \{ 0, 1,2 , 3\}$ is understood. The constants are given by%
\be
  \nonumber
  c^0_n =  0 \, ,
  \quad
  c^1_n =  {1- b_n w_n \ov b_n-w_n}\, ,
  \quad
   c^2_n = i{1+b_n w_n \ov w_n - b_n}  \, ,
  \quad
  c^3_n = {b_n + w_n \ov w_n - b_n } \, ,
\ee
and the $\Sigma_\mu$ are defined by
\be
\nonumber
  \Sigma_0 = \mathbb{1}_{4\times 4}
\, , \quad
\Sigma_1 =
\begin{pmatrix}
0 & 0 & \hskip -0.2cm -1 & 0
\cr
0 & 0 & 0 & \hskip -0.2cm \hskip -0.0cm -1
\cr
-1 & 0 & 0 & 0
\cr
0 & \hskip -0.2cm -1 & 0 & 0
\end{pmatrix}   \, , \quad
\Sigma_2 =
\begin{pmatrix}
0 & i & 0 & 0
\cr
-i & 0 & 0 & 0
\cr
0 & 0 & 0 & -i
\cr
0 & 0 & i & 0
\end{pmatrix}   \, , \quad
\Sigma_3 =
\begin{pmatrix}
0 & 0 & 0 & 1
\cr
0 & 0 & \hskip -0.2cm -1 & 0
\cr
0 & \hskip -0.2cm -1 & 0 & 0
\cr
1 & 0 & 0 & 0
\end{pmatrix}   \, .
\ee
Since the matrices satisfy
\be
   \nonumber
 \Sigma_i \Sigma_j = \delta_{ij} \Sigma_0 + i \varepsilon_{ijk} \Sigma_k \, , 
\ee
they generate an algebra isomorphic to the one generated by $\sigma_0 =\mathbb{1}$ and the three $\sigma_i$  Pauli matrices.
We can thus replace the $\Sigma$ matrices in $\mathcal{R}$ with their $2 \times 2$ counterparts and define
\be
  \nonumber
  \mathcal{R}'  = \prod_{n=N}^1   c^\mu_n \sigma_\mu  \, .
\ee
We have $\mathcal{R}'=\mathbb{1}$ precisely when $\mathcal{R} = \Sigma_0$.
As it happens, $\mathcal{R}'$ can be rewritten in terms of the Lax matrices.
For odd $n$ we have
\be
  \nonumber
    c^\mu_n \sigma_\mu =   i \sigma_3 \Omega_{b_n,w_n}(\zeta=i)\sigma_3 \, ,
\ee
and for even $n$ we write instead the equivalent expression
\be
  \nonumber
    c^\mu_n \sigma_\mu =     -i \sigma_3\Omega^{-1}_{b_n,w_n}(\zeta=i)\sigma_3 \, .
\ee
The $\sigma_3$ matrices and the constant $\pm i$ factors cancel in $ \mathcal{R}'$, which then gives the Lax monodromy from $P$ to $P'$ (sandwiched between two $\sigma_3$ matrices).
Thus, the string closes precisely when the monodromy at $\zeta = i$ (or $x=0$) is trivial, i.e.
\be
  \nonumber
  \Omega(x=0) =   \mathbb{1} \, .
\ee
Notice that the Lax matrix can be written as
\be
\nonumber
  \Omega_{b_2, w_2}^{-1}(x=0) \Omega_{b_1, w_1}(x=0) = Q P^{-1}
\ee
and similarly for the other patches. Thus the monodromy can be written as
\be
  \nonumber
  \Omega(x=0) = Y(2\pi) \, Y^{-1}(0)
\ee
where $Y(\sigma)\in SL(2, \RR)$ is the embedding function on a timeslice; see (\ref{eq:yeq}). If it is periodic, then $\Omega(x=0)$ is trivial.
This criterion generally gives three real constraint equations for the celestial variables.

Recall that for $x=\infty$ or $0$ ($\zeta = 1$ or $i$) the Lax connection yields the left and right connections, respectively, as seen in (\ref{eq:special}).
In the first case we trivially have
\be
   \nonumber
 \Omega(x=\infty) = \mathbb{1} \, ,
\ee
since $\Omega_{b,w}(x=\infty) = \mathbb{1} $ for any values of $b$ and $w$.

\subsection{The spectral curve}

The {\it quasimomentum} $p(\zeta)$ is defined by
\be
 \label{eq:cosp}
   \cos p(\zeta) = \half \tr \Omega(\zeta)
\ee
Since the connection is flat, the result will only depend on $\zeta$ and the homotopy class of the loop. Since the celestial variables are real,  for  $\zeta  \in \RR$ we have $\cos p \in \RR$ with forbidden zones located in the intervals where $|\cos p| > 1$.
Finally, the {\it spectral curve} is defined by the equation
\be
 \nonumber
 \det(\lambda I - \Omega(\zeta) ) = 0
\ee
in the space parametrized by $\{ \lambda, \zeta \} \in \CC^2$.
It is more convenient to consider the {\it algebraic  curve} associated to $p'(x)$ which we write as\footnote{
Henceforth, we will use the spectral parameter $x$ which is related to $\zeta$ via   (\ref{eq:xdef}).}
\be
  \nonumber
  p'(x) = - {{d\ov dx}  \cos p(x) \ov \sqrt{1- \cos^2 p(x)}}  \, .
\ee
Taking the derivative gets rid of the $2\pi n$ phase shift ambiguity in $p$ and the resulting function can be made single-valued on two Riemann sheets.
Properties of this {   curve} can be inferred from the quasimomentum which is known explicitly in terms of the celestial variables  via  (\ref{eq:cosp}).

Let us consider a closed string of $N$ segments as in Figure \ref{fig:spectral}. By inspecting the Lax matrices (\ref{eq:mono}) written in terms of $x$
\be
  \nonumber
\Omega_{b,w}(x) =
{ x \mathbb{1}+\Omega^{(0)}_{b,w} \over  \sqrt{x^2-1}} \,
\qquad\textrm{where}\qquad
\Omega^{(0)}_{b,w} =
 \begin{pmatrix}
-{b+w \over b-w} & {2 b w \over b-w}
\cr
-{2 \over b-w} &  {b+w \over b-w}
\end{pmatrix}
 \, ,
\ee
it is easy to see that the quasimomentum takes the form
\be
  \nonumber
  \cos p(x) = {Q_{N}(x) \ov (x^2-1)^{N \ov 2}} \, ,
\ee
where $Q_{N}$ is a degree $N$ polynomial in $x$ and is a function of the celestial variables $b_i, w_j$ with $i,j = 1,\ldots, N$. The coefficients of the polynomial are conserved charges. Using
\be
  \nonumber
   \lim_{x \to \infty }\cos p(x) = 1 \, ,
\ee
and $\tr  \Omega^{(0)}_{b,w} = 0$, it can also be shown that
\be
   \nonumber
 {d\ov dx} \cos p(x) = {P_{N-1}(x) \ov (x^2-1)^{{N \ov 2}+1}} \, ,
\ee
and
\be
   \nonumber
 1-\cos^2 p(x) = {Q_{2N-2}(x) \ov (x^2-1)^{N}} \, ,
\ee
where $P_{N-1}$ and $Q_{2N-2}$ are degree $N-1$ and $2N-2$ polynomials, respectively.
The spectral curve can then be written as an algebraic curve,
\be
   \nonumber
 p'(x) = \pm { P_{N-1}(x) \ov (x^2-1) \sqrt{Q_{2N-2}(x)}}
\ee
The function can be made single-valued on the hyperelliptic curve defined by $  y^2 =  Q_{2N-2}(x)$. If there is a common factor in $P^2$ and $Q$, then the genus of the curve will be smaller than the naive expectation $g=N-2$. Since the Riemann surface has a finite genus, segmented strings are {\it finite-gap solutions}.

\clearpage

\begin{figure}[h]
\begin{center}
\includegraphics[width=3.5cm]{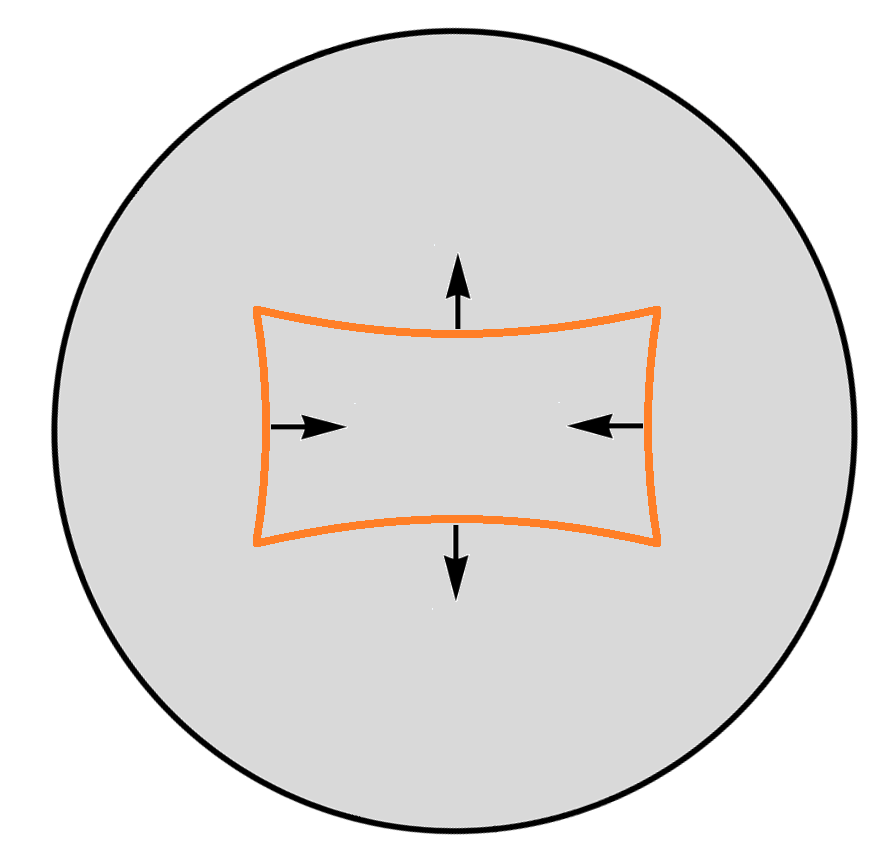} \qquad\qquad
\includegraphics[width=6.6cm]{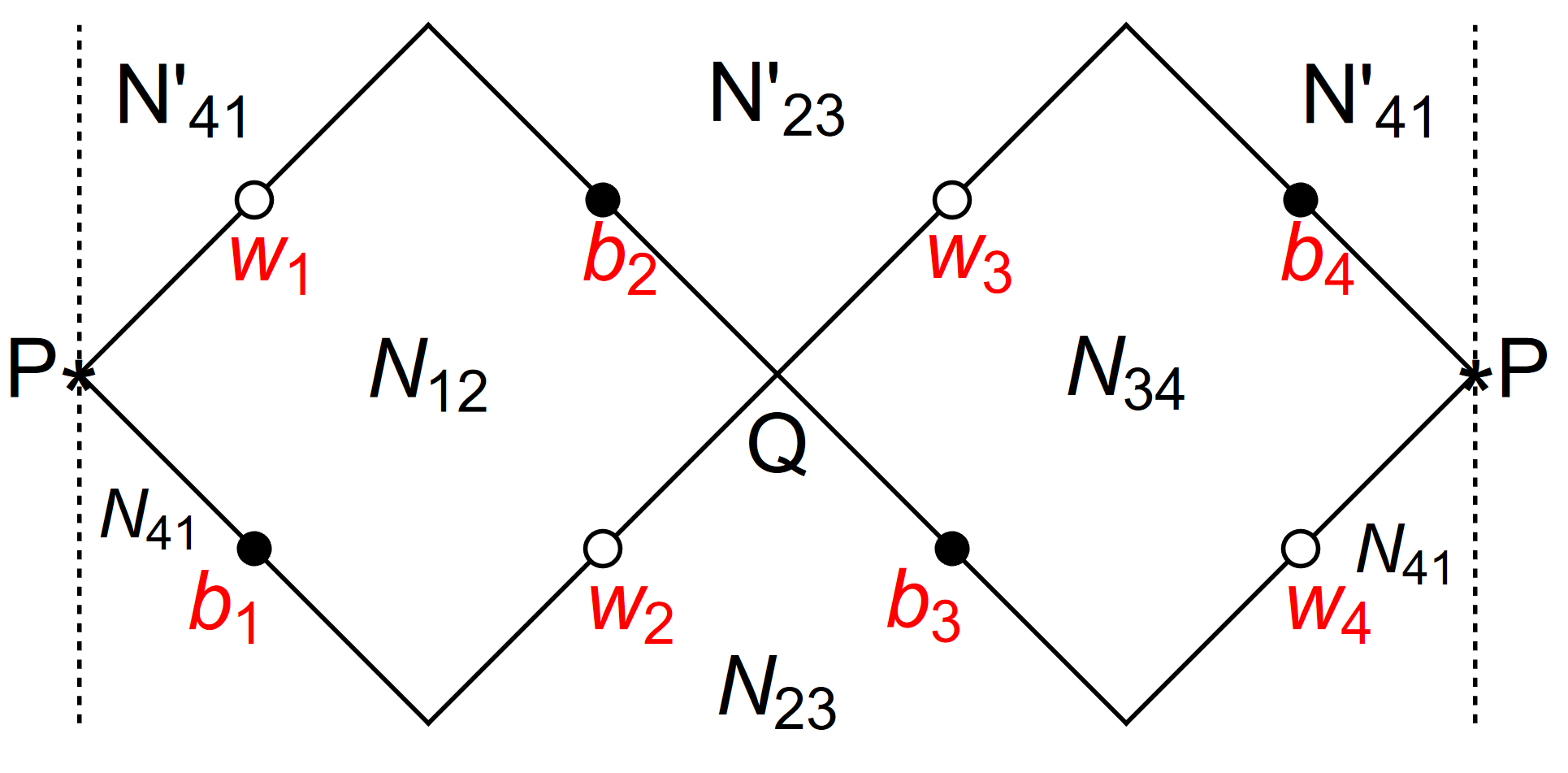}
\caption{\label{fig:4seg} {\it Left:} Segmented string with four kinks on a timeslice of global AdS$_3$. In the applied coordinate system the segments are circular arcs. Arrows indicate the direction of motion of the segments. See \cite{anim} for an animation made by the authors of \cite{Callebaut:2015fsa}. {\it Right:} A part of the worldsheet which should be glued along the dashed lines. $N_{12}, N_{23}, N'_{23}, N_{34}, N_{41}, N'_{41}$ label the normal vectors of the corresponding AdS$_2$ patches. $b_i$ and $w_j$ are celestial variables. $P$ and $Q$ are spacetime points.
}
\end{center}
\end{figure}

\section{An example with four segments}

In this section, as an example, we compute the spectral curve of a closed string with four segments.
The motion of the string has been described in \cite{Callebaut:2015fsa} (see \cite{anim} for an animation made by the authors). The string is seen in Figure \ref{fig:4seg} (left). It is a highest-weight (or primary) state which means that in our global AdS$_3$ coordinate system the `center-of-mass' does not oscillate. After a while neighboring kinks collide and the segments start to move in the opposite direction. The worldsheet with the null kink worldlines are depicted in Figure~\ref{fig:4seg}~(right).

Initial conditions for the kinks were given in \cite{Callebaut:2015fsa}. Let us label them by $i=1,\ldots, 4$. Their initial position is given by
\be
  \nonumber
  h_i = \le( \sqrt{1+r^2}, \, 0, \, r \cos\le( (i-1){\pi\ov 2} + \phi \ri) , \, r \sin\le( (i-1){\pi\ov 2} + \phi \ri) \ri) \in \RR^{2,2} \, ,
\ee
and their velocities are
\be
   \nonumber
 v_i = \le(   0, \, 1, \, (-1)^{i+1} \sin\le( (i-1){\pi\ov 2} + \phi \ri) , \, (-1)^i \sin\le( (i-1){\pi\ov 2} + \phi \ri) \ri) \in \RR^{2,2} \, .
\ee
Here $r$ determines the size of the oscillating string and $\phi$ is an offset angle corresponding to rotations in AdS$_3$.
Note that $h_i^2 = -1$ and $v_i^2 = h_i v_i = 0$ for $i=1,\ldots, 4$.

In order to compute the spectral curve, one needs to obtain the celestial variables $b_i$ and $w_j$ corresponding to the kink velocities.
$v_i$ correspond to the variables $b_1, w_2, b_3, w_4$ which are given by (\ref{eq:aeq}),
\be
  \nonumber
  b_1 = -{\cos \phi \over 1+\sin \phi} \, , \quad
  w_2 = -\cot {\phi \ov 2} \, , \quad
  b_3 = {\cos \phi \over 1-\sin \phi} \, , \quad
  w_4 = \tan {\phi \ov 2} \, .
\ee
The easiest way to calculate the remaining celestial variables is to first compute the normal vectors. These are unit vectors perpendicular to both the  positions of adjacent kinks and their velocities. For instance, we have
\be
 \nonumber
  N_{12}^2 = 1, \, \quad N_{12}\cdot h_1 = 0, \, \quad N_{12}\cdot h_2 = 0, \, \quad N_{12}\cdot v_1 = 0, \,
\ee
which determines the four components of $N_{12}$ up to an overall sign.
We obtain the expressions
\bea
  \nonumber
  N_{12} &=& \le( r, \, -\sqrt{1+r^2}, \, \sqrt{1+r^2}(\cos \phi - \sin \phi), \, \sqrt{1+r^2}(\cos \phi + \sin \phi) \) \, , \\
  \nonumber
  N_{23} &=& \le( r, \, \sqrt{1+r^2}, \, -\sqrt{1+r^2}(\cos \phi + \sin \phi), \, \sqrt{1+r^2}(\cos \phi - \sin \phi) \) \, ,  \\
  \nonumber
  N_{34} &=& \le( r, \, -\sqrt{1+r^2}, \, \sqrt{1+r^2}(\sin \phi - \cos \phi), \, -\sqrt{1+r^2}(\cos \phi + \sin \phi) \) \, ,  \\
  \nonumber
  N_{41} &=& \le( r, \, \sqrt{1+r^2}, \, \sqrt{1+r^2}(\cos \phi + \sin \phi), \, \sqrt{1+r^2}(\sin \phi - \cos \phi) \) \, .
\eea
Here we have fixed the signs of the normal vectors by demanding that the scalar product of adjacent vectors gives plus one. (There is still an overall sign ambiguity, which does not change the results.)
Then, $N'_{23}$ and $N'_{41}$ can be computed from the reflection formula \cite{Vegh:2015ska},
{\footnotesize
\bea
  \nonumber
  && \hskip-0.8cm N'_{23} = -N_{23} + 2 {N_{12} + N_{34} \ov 1+N_{12}\cdot N_{34}} =
 \sqrt{1+r^2} \le( -{r(3+2r^2) \ov (1+2r^2)\sqrt{1+r^2}}, \, {(1-2r^2)\ov 1+2r^2}, \,  \sin \phi+\cos \phi, \, \sin \phi - \cos \phi \) \, , \\
  \nonumber
  && \hskip-0.8cm  N_{41} = -N_{41} + 2 {N_{12} + N_{34} \ov 1+N_{12}\cdot N_{34}} =
  \sqrt{1+r^2}\le(-{r(3+2r^2) \ov (1+2r^2)\sqrt{1+r^2}}, \, {(1-2r^2) \ov 1+2r^2}, \, -  \sin \phi-\cos \phi , \,   \cos \phi-\sin \phi  \) \, .
\eea
}
Now the points $P$ and $Q$ can be determined using the reflection matrix (\ref{eq:reflmat}),
\bea
  \nonumber
  P &=& \mathcal{R}_{b_1, w_4} N_{41}  = \le( \sqrt{1+r^2}, \, r, \, r(\cos\phi + \sin \phi), \, r(\sin\phi-\cos\phi) \ri) \, , \\
  \nonumber
   Q &=& \mathcal{R}_{b_3, w_2} N_{23}  = \le( \sqrt{1+r^2}, \, r, \, -r(\cos\phi + \sin \phi), \, r(\cos\phi-\sin\phi) \ri)\, .
\eea
The missing celestial variables $w_1, b_2, w_3, b_4$ can be computed by solving the equations
\be
  \nonumber
 \mathcal{R}_{b_1, w_1}P = N_{12} \, , \qquad
 \mathcal{R}_{b_2, w_2}Q = N_{12} \, , \qquad
 \mathcal{R}_{b_3, w_3}Q = N_{34} \, , \qquad
 \mathcal{R}_{b_4, w_4}P = N_{34} \, ,
\ee
which gives
\be
  \nonumber
  w_1 = {\sqrt{1+r^2}(1+\cos \phi) + r\sin \phi \ov r(1+\cos\phi) - \sqrt{1+r^2}\sin\phi} \, , \quad
  b_2 = -{2r\sqrt{1+r^2} + (1+2r^2)\cos\phi  \ov 1+(1+2r^2)\sin\phi} \, ,
\ee
\be
  \nonumber
  w_3 = {-2r\sqrt{1+r^2} + (1+2r^2)\sin\phi \ov 1+(1+2r^2)\cos\phi} \, , \quad
  b_4 = {(1+2r^2)\cos\phi-2r\sqrt{1+r^2} \ov 1-(1+2r^2)\sin\phi} \, .
\ee
The monodromy can be computed by
\be
  \nonumber
  \Omega = \Omega_{b_4, w_4}^{-1} \Omega_{b_{3}, w_{3}} \Omega_{b_2, w_2}^{-1} \Omega_{b_1, w_1} \, .
\ee
The explicit expression is too large to be presented here. However, the quasimomentum has a simple form,
\be
 \nonumber
   \cos p(x) = \half \tr \Omega(x) = {1-\le[2+32 r^2(1+r^2)\ri]x^2 + x^4 \ov (1-x^2)^2 } \, ,
\ee
from which we get
\be
  \nonumber
  p' = \pm {1+x^2 \ov 1-x^2} {8r\sqrt{1+r^2} \ov \sqrt{1- [2+16 r^2(1+r^2)] x^2 + x^4 }} \, .
\ee

Finally, we note that conserved quantities such as energy $\Delta$ or angular momentum $S$ in global AdS$_3$ can be obtained by expanding $\Omega$ (or the quasimomentum) near $x=0, \infty$. In the present case $S=0$ therefore we expect to obtain the energy in both cases.

Indeed, near $x=\infty$ one obtains the following simple expression
\be
    \nonumber
  \Omega(x) =  \mathbb{1}- {\Delta \over x}  i\sigma_2  + \mathcal{O}(x^{-2})\, , \qquad \textrm{with} \qquad
  \Delta = 8r\sqrt{1+r^2} \, ,
\ee
where the value of $\Delta$   precisely matches the expression in \cite{Callebaut:2015fsa} (with $2\pi\alpha'=1$).

Near $x=0$ one obtains
\be
    \nonumber
  P^{-1}\Omega(x)P =  \mathbb{1} +  x  i\sigma_2 \Delta + \mathcal{O}(x^{2})\, ,
\ee
where using (\ref{eq:yeq}) we defined the $SL(2)$ matrix $P$
\be
  \nonumber
 P_{a\dot a} =
 \begin{pmatrix}
\sqrt{1+r^2}+r(\sin\phi - \cos\phi)  & r(\sin\phi + \cos\phi-1)
\cr
 r(\sin\phi + \cos\phi+1) &  \sqrt{1+r^2}+r(\cos\phi - \sin\phi)
\end{pmatrix}
 \, ,
\ee
corresponding to the spacetime point of the same name; see Figure~\ref{fig:4seg}~(right).

\clearpage

\section{Discussion}

Segmented strings provide an exact integrable discretization of the string equation of motion. In this paper we have computed the monodromy  of the Lax connection (\ref{eq:monodromy}) on closed segmented strings in AdS$_3$ spacetime. The resulting monodromy matrix is a product of explicit matrices (\ref{eq:mono}) which correspond to the individual string segments. The matrices contain the spectral parameter and the celestial variables which characterize the (null) direction of kinks moving on the string.

Section 8.10 of
\cite{suris} contains Lax matrices for a model which is a one-parameter deformation of the Nambu-Goto string that we have discussed in this paper. The equation of motion is given in $(8.10.14)$ which we repeat here,
\be
  \nonumber
  {h\ov a_{ij} - a_{i,j+1}}+   {h\ov a_{ij} - a_{i,j-1}} =
  {\alpha\ov a_{ij} - a_{i+1,j}}+   {\alpha\ov a_{ij} - a_{i-1,j}}+
  {h-\alpha \ov a_{ij} - a_{i+1,j-1}}+   {h-\alpha \ov a_{ij} - a_{i-1,j+1}} \, .
\ee
If the $\alpha$ deformation parameter is set equal to the $h$ lattice spacing then we recover the string equation of motion (\ref{eq:deqn}). The Lax matrix (\ref{eq:mono}) is identical to the $V_k$ matrix presented in $(8.10.17)$ on page 407 in \cite{suris} if we identify $x_{k+1} = b$ and $x_k = w$ and set $\lambda = {\zeta^2 \ov \zeta^2 -1}$ and $h=\alpha=1$. Note that $\alpha=0$ or $\alpha \to \infty$ also result in the same (\ref{eq:deqn}) equation of motion if a modular transformation is applied on the lattice.

In this paper we used celestial variables appropriate for the \poincare patch. One can switch to global AdS coordinates by replacing $b \to \tan b$ and $w \to \tan w$ \cite{Vegh:2016hwq, Vegh:2019any}.
Note that the string celestial variables need to satisfy extra constraints and therefore the discrete model in the current paper is a reduction of that in \cite{suris}.
These constraints ensure that the patch areas are non-negative and that the string closes in AdS$_3$.
In the case of the Wess-Zumino-Witten model, a further reduction is necessary, which sets certain celestial variables equal; see Figure 3 in \cite{Vegh:2016fcm}.

Note that segmented strings form a dense set in the space of string solutions \cite{Vegh:2021jga} and therefore they are completely generic in AdS$_3$. It would be interesting to see if they can be embedded into top-down constructions of AdS$_3$/CFT$_2$ (see \cite{Babichenko:2009dk, Sfondrini:2014via} and references therein) and study the dual operators in the field theory. It would also be interesting to see if there is any relationship to the discretized `fishchain' string model of \cite{Gromov:2019aku}.

Finally note that in the vertex and AdS$_2$ regions the celestial variables depend on only one lightcone coordinate: $b(z), w(\bar z)$ or $b(\bar z), w(z)$, respectively. The $b$ and $w$ `coordinates' are therefore twisted at points where AdS$_2$ patches meet the vertex patches. This parametrization is somewhat similar to the so-called antimap which performs an untwisting of  brane tilings \cite{Feng:2005gw}; see also \cite{Hanany:2005ss}. In this analogy the black and white vertices of the brane tiling (a.k.a. dimer model) correspond to the vertex and AdS$_2$ regions on the segmented string. Exploring a possible correspondence between brane tilings and segmented strings is left for future work.

\vspace{0.2in}   \centerline{\bf{Acknowledgments}} \vspace{0.2in}
I thank Andrea Cavagli\` a and  Alessandro Torrielli for valuable comments on the manuscript.
The author is supported by the STFC Ernest Rutherford grant ST/P004334/1.

\bibliographystyle{JHEP}
\bibliography{paper}

\end{document}